
\input phyzzx

\def\a{\alpha}
\def\b{\beta}
\def\g{\gamma}
\def\d{\delta}
\def\e{\epsilon}
\def\m{\mu}
\def\n{\nu}

\def\r{\rho}
\def\s{\sigma}
\def\ta{\theta}
\def\tab{\bar{\theta}}
\def\mn{{\mu\nu}}
\def\intdfourp{\int\ {d^4p\over (2\pi)^4}}
\def\intddp{{\int\ {d^D p\over (2\pi)^D}}}
\def\intddpp{{\int\ {d^D p'\over (2\pi)^D}}}
\def\fp{(4\pi)^2}
\def\fpe{(4\pi)^{2-\e/2}}
\def\fpE{(4\pi)^2(\EB/2)^2}

\def\mep{\m^{\e/2}}
\def\dtot#1{{d#1\over #1}}
\def\dToTe#1{{dT\over T^{#1-\e/2}}}
\def\dToT#1{{dT\over T^{#1}}}

\def\tauint{\int_0^{\ \infty} \dtot{T}}
\def\loopTint{\int_0^{\ \infty} \dToT{3}}

\def\loopTiNt#1{\int_0^{\ \infty} \dToT{#1}}
\def\loopTiNte#1{\int_0^{\ \infty} \dToTe{#1}}
\def\Caa{{(ig)^N\over\fpE}}

\def\Cabe{{(ig\mep)^N\over\fpe}}
\def\Cb{-4\ {(ig)^N\over2\fp}}
\def\Cc{{(ig)^N\over 2\fp}}

\def\taupint{\int_0^{\ T} d\tau}
\def\tauppint{\int_0^{\ T} d\tau'}
\def\tint#1#2{\int_0^{t_{#2}} dt_{#1}}
\def\tNint{\int_0^T dt_{N-1}}
\def\xdt{\dot x}
\def\Gdt{\dot G}
\def\Gdd{\ddot G}
\def\Gf#1{G_F^{#1}}
\def\GFpp{G_F^{++}}
\def\GFpm{G_F^{+-}}
\def\GFmp{G_F^{-+}}
\def\GFmm{G_F^{--}}
\def\GFp{G_F^{(0)}}
\def\GFa{G_F^{(\half)}}
\def\Gbd#1{\Gdt_B^{#1}}
\def\Gbdd#1{\Gdd_B^{#1}}
\def\expkk{\exp\big[\sum_{r<s}k_r\cdot k_s G_B^{sr}\big]}
\def\kk#1#2{k_#1\cdot k_#2}
\def\ek#1#2{\e_#1\cdot k_#2}
\def\ke#1#2{k_#1\cdot\e_#2}
\def\ee#1#2{\e_#1\cdot\e_#2}
\def\Dp{{\cal D}p}
\def\Dx{{\cal D}x}

\def\Dphi{{\cal D}\phi}
\def\Dpsi{{\cal D}\psi}
\def\fourth{{1\over 4}}
\def\Tr{{\rm Tr}}
\def\sign{{\rm sign}}
\def\ch{{\rm cosh}}
\def\sh{{\rm sinh}}
\def\linv{}

\def\Eotu{{\EB\over 2}}
\def\otuE{{1\over 2\EB}}
\def\rarr{\rightarrow}
\def\half{{1\over 2}}
\def\del{\partial}
\def\dbyd#1{{\del\over \del #1}}

\def\bar#1{\overline{#1}}
\def\dslash{\not{\hbox{\kern-2pt $\partial$}}}
\def\eslash{\not{\hbox{\kern-2pt $\e$}}}
\def\Dslash{\not{\hbox{\kern-4pt $D$}}}
\def\Qslash{\not{\hbox{\kern-4pt $Q$}}}
\def\pslash{\not{\hbox{\kern-2.3pt $p$}}}
\def\kslash{\not{\hbox{\kern-2.3pt $k$}}}
\def\qslash{\not{\hbox{\kern-2.3pt $q$}}}
\def\psm{\psi_{-}}
\def\psp{\psi_{+}}
\def\EB{{\cal E}}
\def\Lgr{{\cal L}}
\def\cnst{{\cal N}\ }
\def\O{{\cal O}}
\def\V{{\cal V}}
\def\BK{Bern-Kosower}
\def\BnK{Bern and Kosower}
\def\rem#1{}
 \def\sqr#1#2{{\vcenter{\hrule height.#2pt                       %
  \hbox{\vrule width.#2pt height#1pt \kern#1pt \vrule width.#2pt}%
   \hrule height.#2pt}}}                                         %
 \def\square{\mathchoice\sqr64\sqr64\sqr63\sqr63}                %

\def\JETPL{JETP~Lett.}
\def\NC{Nuovo~Cimento}

\def\NPB{Nucl.~Phys.~B}

\def\PLB{Phys.~Lett.~B}
\def\PRD{Phys.~Rev.~D}
\def\PRL{Phys.~Rev.~Lett.}
\def\Fpre{Fermilab preprint}
\def\Ppre{Pittsburgh preprint}

\def\ZBnDAK{Z.~Bern and D.~A.~Kosower}

\pubnum{5757}
\date{February, 1992}
\pubtype{(T)}
\titlepage

\title{Field~Theory~Without~Feynman~Diagrams:
One-Loop~Effective~Actions}
\author{Matthew J. Strassler\foot{Work supported in part by an NSF
Graduate Fellowship and by the Department of Energy, contract
DE-AC03-76SF00515.}}
\SLAC

\centerline{}
\centerline{In memory of Brian J.~Warr. }
\abstract{ In this paper the connection between standard perturbation
theory techniques and the new Bern-Kosower calculational
rules for gauge theory is clarified.  For one-loop effective actions
of scalars, Dirac spinors, and vector bosons
in a background  gauge field, \BK-type rules are derived
without the use of either string theory or Feynman diagrams.  The
effective action  is written as a one-dimensional path integral, which
 can be calculated to any order in the gauge coupling;
evaluation leads to Feynman parameter integrals directly,
bypassing  the usual algebra required from Feynman diagrams, and
leading to compact and organized expressions.  This formalism
is valid off-shell,  is explicitly gauge invariant, and
can be extended to a number of other field theories.}

\submit{Nuclear Physics B}

\endpage

\chapter{Introduction}

\REF\treelevl{D.~A.~Kosower, B.-H.~Lee and V.~P.~Nair,
\PLB201 (1988) 85;
D.~A.~Kosower, \NPB335 (1990) 23; \NPB315 (1989) 391}
\REF\MPX{M.~Mangano and S.~J.~Parke, \NPB299 (1988) 673;
M.~Mangano, S.~Parke and Z.~Xu, \NPB298 (1988) 653}
\REF\BG{F.~A.~Berends and W.~T.~Giele,
 \NPB306 (1988) 759; B294 (1987) 700}
\REF\XZC{Z.~Xu, D.~Zhang and L.~Chang, \NPB291 (1987) 392;
Tsinghua University preprint TUTP-84/3 (1984), unpublished}
\REF\ElSext{R.~K.~Ellis and J.~C.~Sexton, \NPB269 (1986) 445}
\REF\BKrules{\ZBnDAK, \Fpre\ FERMILAB-PUB-91-111-T (1991)}
\REF\BKPascos{\ZBnDAK, \Ppre\  PITT-91-03 (1991)
Presented at PASCOS '91 Conf., Boston, MA, USA, 1991.}
\REF\BKcolor{\ZBnDAK, \NPB362 (1991) 389}
\REF\BKother{\ZBnDAK,
Proc. Polarized Collider Wkshp., University Park, Pa., USA, 1990
(American Institute of Physics, New York, 1991) p.~358;
\PRL\ 66         (1991) 1669;
Proc. Perspectives in String Thy., Copenhagen, Denmark, 1987
(World Scientific, Singapore, 1988) p.~390;
\NPB321 (1989) 605;  \PRD38 (1988) 1888;
Z.~Bern, D.~A.~Kosower and K.~Roland,  \NPB334 (1990) 309}
\REF\BDmap{Z.~Bern and D.~C.~Dunbar, \Ppre\ PITT-91-17  (1991)}
\REF\Feynman{R.~P.~Feynman, Phys. Rev. 84 (1951) 108;
Phys. Rev. 80 (1950) 440}
\REF\Swing{J.~Schwinger, Phys. Rev. 82 (1951) 664}
\REF\GSW{M.~B.~Green, J.~H.~Schwarz and E.~Witten, Superstring Theory
(Cambridge Univ. Press, Cambridge, 1987)}
\REF\BdVH{L.~Brink, P.~Di Vecchia and P.~Howe,  \NPB118 (1977) 76}
\REF\NAbCase{A.~Barducci, R.~Casalbuoni and L.~Lusanna,
\NPB124 (1977) 93;
A.P.~Balachandran , P.~Salomonson,
B.~Skagerstam and J.~Winnberg,
\PRD15 (1977) 2308}
\REF\Berez{F.~A.~Berezin and M.~S.~Marinov, \JETPL\ 21 (1975) 320}
\REF\Casspin{R.~Casalbuoni,  \NC\ 33A (1976)  389}
\REF\Cascolor{A.~Barducci, F.~Buccella,
R.~Casalbuoni, L.~Lusanna and E.~Sorace,   \PLB67 (1977) 34}
\REF\ThankLD{L.~J.~Dixon, private communication}
\REF\Abbott{L.~F.~Abbott,  \NPB185 (1981) 189}
\REF\GSO{F.~Gliozzi, J.~Scherk and D.~Olive, \NPB122 (1977) 253}
\REF\West{P. West, Introduction to Supersymmetry and Supergravity
(World Scientific, Singapore, 1990) and references therein}
\REF\ChanPat{J.~E.~Paton and H.~M.~Chan, \NPB10 (1969) 516}
\REF\thooft{G.~'t~Hooft, \NPB72 (1974) 461}
\REF\tHgauge{G.~'t~Hooft, \NPB35 (1971) 167}
\REF\GJ{J.-L.~Gervais and A.~Jevicki, \NPB110 (1976) 93}
\REF\backstring{
C.~G.~Callan, E.~J.~Martinec, M.~J.~Perry and D.~Friedan,
 \NPB262 (1985) 593;
E.~S.~Fradkin and A.~A.~Tseytlin,
\NPB261 (1985) 1;  \PLB163 (1985) 123;  \PLB160 (1985) 69;
A.~Abouelsaood, C.~G.~Callan, C.~R.~Nappi and S.~A.~Yost,
\NPB280 (1987) 599}

In the past year significant advances have been made in techniques
for calculating one-loop scattering amplitudes in gauge theories.
Following on the successes of several authors
at applying string theory and various technical innovations
to tree-level gauge theory
calculations\refmark{\treelevl,\MPX,\BG,\XZC},
Z.~Bern and D.~A.~Kosower have
derived new rules from string theory
for one-loop gauge theory scattering
amplitudes.
In reference \BKrules, they present the derivation of the rules
and apply them to the computation of
two-to-two gluon scattering at one loop, which previously
was difficult enough to challenge the
most expert calculators.\refmark\ElSext\  In reference \BKPascos,
they    present their rules in a compact form and work
a simple example.  Although obtained  from string theory,
the \BK\ rules do not refer to string
theory in any way, but as they also bear little resemblance to
Feynman rules, it is of interest to derive them
directly from field theory.  Bern and Dunbar\refmark\BDmap
showed how to map the \BK\ rules onto Feynman diagrams and
demonstrated that the background field method plays an important role;
in this paper I take the opposite route, deriving \BK\ rules
from the field theory path integral with the use of the background
field method.

The main result of this paper is that calculational rules
similar to those of \BnK\ can be derived
from first-quantized field theory.  Unlike the
``connect-the-dots'' approach of Feynman diagrams,
first-quantized field theory (particle theory) views
a particle in a loop as a single entity, acted on by
operators representing the effects of external fields.
We are all well-accustomed to this approach in atomic physics,
where electromagnetic fields are treated as operators
acting on quantum mechanical electrons, but to my
knowledge it rarely been used for calculations with
relativistic particles.  (Feynman presented formulas
similar to those discussed in this paper but did not use them to
develop perturbation theory.\refmark\Feynman)
In any case, it will not surprise those
familiar with first-quantized strings
that just as string theory amplitudes are evaluated as
two-dimensional path integrals, so particle theory
amplitudes can be calculated using one-dimensional path
integrals --- the path integrals of quantum mechanics.

In this paper I address the issue of the effective action
at one loop.  In section 2, I construct the one-loop
effective action of  a scalar particle in
a background gauge field, and derive rules almost identical to
those of \BnK.  In sections 3 and 4 I generalize this approach
to Dirac spinors and vector bosons.
  Section 5 contains a study of the
integration-by-parts procedure involved in the \BK\ rules, and
an illustration of its relation to manifest gauge invariance.
After a short comment (section 6)
on an alternative organization of color traces
in this formalism, I conclude in section 7
with some extensions of this approach  to other field theories.

\chapter{The Effective Action of a Scalar in a Background Field}

In this section, I will show that the one-loop
effective action of a particle
in a background field, when written as a one-dimensional path
integral, is calculable at any order in the coupling constant $g$.
A particle in a loop    can be described as a
simple quantum mechanical system existing
for a finite, periodic time, or, alternatively,
as a one-dimensional field theory on a compact space;
external fields act as operators on the particle Hilbert space,
just as in usual quantum mechanics.
At any order in the external field, the effective action
is a correlation function of these operators in a free and therefore
soluble theory, and can be expressed in a compact form.
By writing the effective action
as a one- rather than a four-dimensional path integral
I employ quantum mechanics instead of quantum field theory;
as string theory in its present form is a first-quantized
theory, it is not especially surprising that
the expressions
found from string theory by \BnK\
are of the same form as those found in this paper.

Working initially in Euclidean spacetime,
let us first consider the one-loop vacuum energy       of a free
scalar field, with Lagrangian
$$ \Lgr = -(\del\phi^{\dag})\cdot(\del\phi) -  m^2\phi^{\dag}\phi \ .
  \eqn\freeact  $$
First represent it in terms of Schwinger
proper time $\tau$:\refmark{\BKrules,\Swing}
$$ \eqalign {
\log Z = & \log \Big[ \int \Dphi\ e^{-\int d^4x \Lgr}\Big]
= - \log \big[\det(-\del^2+m^2)\big] \cr
= &- \Tr \log (-\del^2+m^2)
 = \tauint \intdfourp\  \exp\big[-\half\EB T (p^2+m^2)\big] \ .
\cr } \eqn\introtau $$
The parameter $\EB$ (the einbein) is an arbitrary constant.
Next convert this result into a path integral over $x^\m(\tau)$:
$$ \eqalign{ \log Z
= & \tauint \int\ \Dp\ \Dx\  \exp\big[\taupint\ ip\cdot\xdt\ \big]
 \  \exp[-\half\EB\taupint (p(\tau)^2+m^2)\ ]
\cr
= &\ \tauint\ \cnst \int_{x(T)=x(0)}
 \Dx\  \exp\big[-\taupint ({\otuE}\xdt^2\ +{\EB\over 2} m^2)\big]
\ ,\cr }
\eqn\freepart $$
where the normalization constant $\cnst$ is
$$ \cnst = \int\Dp\ e^{-\half\taupint \EB p^2}
                                       \eqn\ddpint $$
and satisfies
$$ \cnst\int\Dx\ e^{-\taupint\otuE\xdt^2}
= \intddp\ e^{-\half\EB Tp^2}
= [2\pi \EB T]^{-D/2} \ .   \eqn\dpdxcnst $$

The result of \freepart\ is a one-dimensional field theory:
the particle position $x^{\mu}(\tau)$ is a set of
four fields living in the
one-dimensional space of proper time, called the worldline.
Eq.~\freepart\ contains
the well-known first-order form of the action for a
free particle\refmark\GSW, which,
unlike the usual Einstein action,
is well defined in the massless limit:
$$ L = \otuE\xdt^2 \  . \eqn\massless $$
(Since a massless particle has no internal clock,
$\tau$ is not actually proper time in this case, though I will loosely
continue to refer to it as such.)
Classically, the action is reparametrization
invariant (that is, invariant under $\tau\rarr\tau'(\tau))$
when the
einbein, the square root of the one-dimensional metric, is
chosen to transform in the proper way.  On the other hand, the
functional integral in \freepart\ is not invariant unless one
integrates over the einbein as well.  In the present work
I will keep $\EB$ constant
and ignore the reparametrization invariance,
since it is not needed for practical
results.

Now let us consider the same system (massless, for simplicity)
in a classical background Abelian gauge field $A_{\mu}(x)$:
$$ \Lgr = \phi^{\dag} D^2\phi       \eqn\backact  $$
where $D_{\mu} = \del_{\mu} - igA_{\mu}$.
The object of interest is the one-loop
effective action generated by \backact,
 as a function of $A_{\mu}$.
In analogy to eqs.~\introtau--\freepart,
$$ \eqalign{\Gamma[A] = &-\log\ [\det\ (-D^2)\ ] \cr
 = & +\tauint\ \intdfourp\
 \bra{p}\exp[-\half\EB T (p+gA(x))^2]\ket{p}
\cr   = &\tauint\ \cnst \int \Dx\
\exp\Big[-\taupint (\otuE\xdt^2\
+\ igA[x(\tau)]\cdot\xdt)\Big] \ . \cr
}
\eqn\effactSP $$
Continuing this result to Minkowski spacetime
and redefining $\EB\rarr -\EB$
gives
$$ \eqalign{\Gamma[A] =
 &\tauint\ \cnst \int \Dx\
\exp\Big[-\taupint (\otuE\xdt^2\
-\ igA[x(\tau)]\cdot\xdt)\Big] \cr
 = & \tauint \cnst \int \Dx\
e^{-\taupint (\otuE\xdt^2)}\ \exp[ig \oint dx\cdot A(x)]\ .
\cr }
\eqn\EASPMink $$
This expression
is immediately recognizable as the expectation value of
a Wilson loop of the background field, in a certain ensemble of loops.
It is therefore explicitly
gauge invariant with respect to the background gauge field,
 as it should be.

  The non-Abelian generalization of this structure is
easy to guess;
one merely inserts a trace over color states:
$$ \Gamma[A]  = \tauint\ \cnst \int \Dx\
\linv\Tr_{\rm R} \exp\Big[-\taupint (\otuE
\xdt^2\ -\ igA[x(\tau)]\cdot\xdt)\Big]\ ,
\eqn\effactSG $$
where the gauge field is a matrix $A^a_\m T^a$ in the gauge group
representation R of the scalar.
 Notice that the usual path-ordering in the Wilson
loop appears here as proper-time--ordering, implicit
in the path integral construction.

Let us now consider the expansion of this effective action to order
$g^N$, which is equivalent to studying the one-particle-irreducible
(1PI)   Feynman
diagrams with $N$ background gluons and one scalar loop.
(By ``gluon'' I mean any
non-abelian vector boson.)
  In the standard Feynman graph technique
  there are a number of such diagrams, involving both
the one-gluon/two-scalar vertex 
and the two-gluon/two-scalar vertex.  
Here, there is only {\it one}
 computation.  We expand the
Wilson loop to order $g^N$:
$$ \Gamma_N[A] =  {(ig)^N\over N!}\tauint\ \cnst\int \Dx\
 e^{-\taupint \otuE\xdt^2}
 \linv \Tr \Big(\prod_{i=1}^{N}
  \int_0^{\ T} dt_i A[x(t_i)]\cdot\xdt(t_i)\Big) \ .
\eqn\EAorderN $$
Up to this point the background field is completely arbitrary.
To compute $\Gamma_N[A]$ as a function of momentum eigenstates,
we insert for  $A_\m$ a sum of classical modes of definite (outgoing)
momentum $k_i$, polarization $\e_i$, and gauge charge $T^{a_i}$:
$$A^\m(x) = \sum_{i=1}^N
T^{a_i} \e_i^{\mu} e^{ik_i\cdot x}
\eqn\Astate $$
Again $T^{a_i}$ is a matrix in the representation of the scalar.
Inserting this function into \EAorderN\ and keeping only
the terms
in which each mode appears precisely once, we find:
$$ \eqalign {
\Gamma_N(k_1,\dots,k_N) = (ig)^N \tauint\ \cnst &\int \Dx\
 e^{-\taupint \otuE\xdt^2} \cr
 \linv \Tr(T^{a_N}\dots&  T^{a_1}) \prod_{i=1}^{N}
  \int_0^{\ t_{i+1}} dt_i\ \e_i\cdot\xdt(t_i) e^{ik_i\cdot x(t_i)}
\cr}
\eqn\EAwVOs $$
   plus terms with all other
orderings of the $t_i$ and $T^{a_i}$.
(Here $t_{N+1} \equiv T$.)
Notice that for a given integration ordering (= path-ordering around
the loop = proper-time-ordering = color-trace-ordering),
the color information factors out.
\rem{In string theory
the color trace is known as a Chan-Paton factor.\refmark\ChanPat\
The utility of computing color-ordered tree-level partial amplitudes
using color-ordered Feynman
diagrams was emphasized by
Mangano, Parke and Xu\refmark\MPX; a study of color-ordering
in loop graphs was performed by \BnK.\refmark\BKcolor\  }
For pure vector field backgrounds, only one color-ordering is
actually necessary, as all other orderings are related to it by
permutation of labels; because of this,
I will consider for the remainder of this paper
only one color ordering
at a time, leaving the sum over color orderings implicit.

String theorists will immediately recognize eq.~\EAwVOs;
the string
theory version of this formula
gives the expectation value of $N$ ``vertex operators'',
which in string theory can be
interpreted as a {\it scattering amplitude} of $N$ strings.  For
strings, duality of the $s$ and $t$ channels implies
that not only the
one-particle-irreducible
loop  but also the {\it trees} which are
sewn onto the loop are calculated in this way.  In particle theory,
however,
eq.~\EAwVOs\
computes {\it only the effective action}, the
one-particle-irreducible
graphs with a scalar loop, at order $g^N$.
Still, it has the advantage of being well-defined even for
off-shell external gauge fields,
unlike usual string theory.

To calculate this expectation value I use the standard
path integral methods of string perturbation theory.\refmark\GSW\
First, disregard the polarization vectors, and notice
that the momenta $k_i$ in \EAwVOs\ serve as sources
of the four fields $x^\mu(\tau)$:
$$ J^\mu(\tau) =  \sum_{j=1}^N\ i\ k_j^\mu \delta(\tau-t_j)
\eqn\source $$
       Using eq.~\dpdxcnst, we find
$$ \eqalign { \Gamma_N&(k_1,\dots,k_N)    \cr
 = &\Caa \linv \Tr(T^{a_N}\dots T^{a_1})
\loopTint  \prod_{i=1}^{N}  \int_0^{\ t_{i+1}} dt_i\  \cr
& \ \ \ \ \ \ \ \ \ \ \exp[\taupint \tauppint \bigg( - \half
J^\mu(\tau)G_B(\tau,\tau')J_\mu(\tau')\bigg)] \cr
 = &\Caa \linv \Tr(T^{a_N}\dots T^{a_1})
\loopTint  \prod_{i=1}^{N} \Big( \int_0^{\ t_{i+1}} dt_i\ \Big) \cr
& \ \ \ \ \ \ \ \ \ \
\exp\Big[\sum_{i,j=1}^N \half k_i\cdot k_j G_B(t_i,t_j)\Big] \ . \cr}
\eqn\EAwJGJ $$
Here $G_B(t,t')$ is the one-dimensional propagator on a loop, which I
will discuss later.  (The $B$ indicates that $G_B$ is the
Green function of the Bosonic field $x^\m$.)

The standard method for including the polarization
vectors is to exponentiate them, with the understanding that the
only terms to be used are those which contain one $\e_i$:
$$\xdt\cdot A_i^{\mu}(x(t_i)) =
T^{a_i}  \exp\big[\e_i\cdot\del_{t_i}x(t_i) + ik_i\cdot x(t_i)\big]
\Big|_{{\rm linear\ in\ \e_i}}
\eqn\Anew $$
This leads  to a new source for $x^\m$:
$$ J^\mu(\tau) = \sum_1^N\ \delta(\tau-t_i)\
(\e_i^\mu \del_{t_i} + ik_i^\mu)  \ .
\eqn\newsource $$
Integration over $x(\tau)$ gives
$$ \eqalign {
 \Gamma_N(k_1,\dots,k_N)
  = \Caa \linv &\Tr(T^{a_N}\dots T^{a_1})
\loopTint  \Big(\prod_{i=1}^{N} \int_0^{\ t_{i+1}} dt_i\ \Big) \cr
\exp\Big[\half \sum_{i,j=1}^N &\big(k_i\cdot k_j G_B(t_j-t_i)] \cr
&-\ 2i k_i\cdot \e_j \dbyd{t_j} G_B(t_j-t_i) \cr
& -\ \e_i\cdot \e_j {\del^2\over\del t_i\del t_j}  G_B(t_j-t_i)
\big)\Big]  \Big|_{\rm linear\ in\ each\ \e}\ ;
\cr}
\eqn\EAwGs $$
again only terms in which each polarization vector
appears exactly once are to be used.
String theorists and those familiar with the work of
Bern and Kosower\refmark\BKrules\
will recognize this form for the amplitude.

Now let us study the Green function (one-dimensional propagator),
which satisfies the equation
$$ {1\over\EB} \del_t^2 G_B(t,t') =  \delta(t-t') \
\eqn\GFeqn $$
with appropriate boundary conditions.
If we were studying this Green function on the real line, the
solution would be
$$ G_B(t,t') = \Eotu |t-t'|\  +\ A \ +\ Bt \ . \eqn\GFline $$
Notice that the Green function is finite as $t$
approaches $t'$, which is not true for higher dimensions;
thus there are no operator singularities when $x$ fields come
together.  This naturally simplifies many discussions.

To find the Green function on a circle of
circumference $T$, one must first note that
eq.~\GFeqn\ has no solution on the loop; it is equivalent to solving
Poisson's equation for a charge
in a compact space, for which the potential is infinite
unless there is a background
charge that makes the total space neutral.
Since we have one unit of charge at $t'$, we should
add a uniform background charge of density $-1/T$.
The new Green function equation is
$$ {1\over\EB}\del^2 G_B(t,t') =  \delta(t-t') - {1\over T}\ ,\
\eqn\GFdd $$
which has a solution when the condition of periodicity
in $t\rarr t+T$ is imposed:
$$ G_B(t,t') = \Eotu\Big(|t-t'| - (t-t')^2/T   \Big)\
+\ {\rm constant}\ . \eqn\GFloop $$
It is convenient to take the arbitrary constant
to be zero, as any additive
constant in $G_B$ cancels out of eq.~\EAwGs.
This function has as its derivative
$$ \del_t G_B(t,t') = \Eotu\big(\sign(t-t') - 2(t-t')/T\big)\  ,
\eqn\GFdloop $$
and its second derivative is given in eq.~\GFdd.  Note that
$G_B$ and $\del_t^2 G_B$ are symmetric in their arguments,
while $\del_t G_B$ is antisymmetric.
 These functions
(up to a multiplicative constant)
were found by Bern and Kosower\refmark\BKrules\
from the  one-loop
string theory bosonic
Green function and its derivatives, in the limit where
$t-t'$ is large compared
to the width of the string theory torus.
Roughly adhering to their conventions,
I shall use the notation $G_B^{ji}\equiv G_B(t_j-t_i)$,
$\Gbd{ji}\equiv\del_{t_j} G_B^{ji}$, and
$\Gbdd{ji}\equiv\del^2_{t_j}G_B^{ji}$.

 It is useful to transform
eq.~\EAwGs\ into a simpler form.  First,
through the use of the crucial relations
$G_B(t,t)=0$ and (by antisymmetry in $t$ and $t'$)
$\del_tG_B(t,t)\equiv 0$, the terms in \EAwGs\
with $\e_i\cdot k_i$ and $k_i^2$
are removed {\it without} the use of on-shell
conditions.
Second, it is useful to replace $t_i\rarr u_iT$, where $u_i$
is dimensionless; $N$ powers of $T$ are thereby factored out.
Next, observe that the integral over $u_N$ is trivial; after the first
$N-1$ integrals no dependence on the $u_i$ remains, and so the last
integral, which contributes a factor of unity,
can be dropped. It is useful to choose the origin of proper time
 by fixing $t_N\equiv T$, and as a consequence we should
 sum only over color traces which are
not related by cyclic permutation.
A further advantage is gained by
choosing the (dimensionless) gauge $\EB=2$.
Lastly, anticipating the use of dimensional regularization, I
redo the integral over momentum in $4-\e$ dimensions as in
eq.~\dpdxcnst, with $\mu$ the arbitrary mass parameter.
(For the remainder of this paper,
the conventions chosen above will
be used except where explicitly noted.)

The result of all these changes is
$$ \eqalign {
 \Gamma_N(k_1,\dots,k_N) =&
\Cabe \linv \Tr(T^{a_N}\dots T^{a_1}) \cr
& \loopTiNte{3-N}  \int_0^{1}         du_{N-1}
  \int_0^{\ u_{N-1}} du_{N-2}\ \cdots
  \int_0^{\ u_{2}} du_1 \cr
\exp\Big[&\sum_{i<j=1}^N k_i\cdot k_j G_B^{ji}\Big] \cr
\exp\Big[&\sum_{i<j=1}^N \big(
-\ i(k_i\cdot \e_j -
     k_j\cdot \e_i)\ \Gbd{ji}
 +\ \e_i\cdot \e_j\ \Gbdd{ji}\big)\Big]\
  \Big|_{\rm linear\ in\ each\ \e}\ ;
\cr}
\eqn\EAasBK $$
plus all other proper-time-orderings.
Meanwhile the Green functions have become
$$\eqalign{
  G_B(t,t') &\equiv T\Big(|u-u'| - (u-u')^2 \Big)\ ; \cr
 \del_t G_B(t,t') &\equiv \big(\sign(u-u') - 2(u-u')\big)\  ; \cr
 \del_t^2 G_B(t,t') &\equiv {2\over T}\big(\delta(u-u') - 1\big)
\  .\  \cr}    \eqn\GBdefn $$
Comparison with reference \BKrules\ or \BKPascos\ shows that
the correspondence between eqs.~\EAasBK\ and \GBdefn\
and the Bern-Kosower rules for the
one-particle-irreducible scalar loop
diagram with $N$ gluons is exact, up to differences in conventions.

Following \BnK\refmark\BKPascos, let us study the result of \EAasBK.
The overall constant factor, the color trace and the integrals are
easy to understand.  The exponential
$$ \exp\Big[\sum_{i<j=1}^N k_i\cdot k_j G_B^{ji}\Big]
= \exp\Big[T\sum_{i<j=1}^N k_i\cdot k_j
  \Big(|u_j-u_i| - (u_j-u_i)^2 \Big)\Big]  \eqn\GBexpo $$
is a ubiquitous factor which, after the integration over $T$, becomes
the usual
Feynman-parameterized
denominator for a scalar loop
integral (notice it contains no polarization vectors, and is
thus spin-independent):
$$ \int_0^\infty\ dT\ T^\a
 \exp\Big[\sum_{i<j=1}^N k_i\cdot k_j G_B^{ji}\Big]
= {\Gamma(\a+1)\over\Big[-\sum_{i<j=1}^N k_i\cdot k_j
  \Big(|u_j-u_i| - (u_j-u_i)^2 \Big)\Big]^{\a+1}} \ . \eqn\GBTint $$
The remaining term,
$$ \exp\Big[\sum_{i<j=1}^N \big(
-\ i(k_i\cdot \e_j -
     k_j\cdot \e_i)\ \Gbd{ji}
 +\ \e_i\cdot \e_j\ \Gbdd{ji}\big)\Big]\
  \Big|_{\rm linear\ in\ each\ \e}\ ,     \eqn\gkfdefn    $$
which I shall call the ``generating kinematic factor'',
provides the numerator of the Feynman parameter integral.
It is the only
part of \EAasBK\ (other than the overall normalization) which has
any information about the type of particle in the loop or the
nature of the external field.
It is also the only part of the result
which cannot be guessed on general grounds;
we undergo the usual
struggles with Feynman diagrams and loop momentum integrals in order
to obtain precisely this piece of information.

However, the form of the
generating kinematic factor causes some practical problems.
At first glance \EAasBK\
appears to have expressed the entire result in such a way
that one has exactly one set of
Feynman parameter integrals for each color
trace, but this is not quite true.  The difficulties
stem from the $\Gdd_B$ functions.  The first problem is
that each term with M $\Gdd_B$'s has M fewer powers of $T$
than terms without $\Gdd_B$'s,
so a number of different integrals over $T$ must be performed.
The second problem is that hiding inside each $\Gbdd{ji}$
is a delta function in $t_j-t_i$.
The evaluation of this delta function
gives the contribution of the Feynman diagram in which gluons $i$
and $j$ come onto the loop {\it via} a four-point vertex.
Thus the expression in eq.~\EAasBK\ contains all of the 1PI
Feynman diagrams, in fact, and each one
generates slightly different integrals and integrands.
(Fortunately, these problems can be dealt with\refmark\BKPascos,
as I will discuss in section 5.)

There is a subtle factor of two concerning the delta function
in $\Gbdd{ij}$.
\rem{Consider smoothing out the singularity slightly;
then, in order to maintain the symmetries of $G_B$ and its derivatives
only half of the delta function actually contributes to
a given color trace.  In other words,
one must be careful to assign half of the delta function in
$\Gbdd{ij}$ to
$\Tr(\cdots T^{a_i} T^{a_j} \cdots)$ and the other half to
$\Tr(\cdots T^{a_j} T^{a_i} \cdots)$. }
Consider smoothing out the singularity slightly;
then, in order to maintain the symmetries of $G_B$ and its derivatives
one must assign half of the delta function to
$t_i>t_j$ and the other half to $t_i<t_j$.
In other words,
   the delta function  is split  between
$\Tr(\cdots T^{a_i} T^{a_j} \cdots)$ and
$\Tr(\cdots T^{a_j} T^{a_i} \cdots)$.

I now present the simplest possible example, the contribution
of a massless scalar
to the gluon vacuum polarization.  There are two Feynman diagrams,
the first of which involves two three-point vertices, the other of
which involves a single four-point vertex.  The former is given
by
$$ (ig)^2 \Tr(T^aT^b) \intddp
 {(i)^2\e_1\cdot (2p-k_1) \e_2\cdot(2p-k_1) \over p^2(p-k_1)^2}
  \eqn\diagrama $$
where $k_1$ is the momentum flowing out along gluon 1.
The second diagram is given by
$$ 2ig^2 \Tr(T^aT^b) \intddp
 {i\e_1\cdot\e_2 \over p^2} \ .
  \eqn\diagramb $$
I now use the Schwinger trick\refmark\Swing\
to evaluate \diagrama\ in a form conducive to comparison with
the expression in \EAasBK.
$$  \eqalign{ \intddp
 &{-\e_1\cdot (2p-k_1) \e_2\cdot(2p-k_1) \over p^2(p-k_1)^2} \cr
 = & \int_0^\infty T dT \int_0^1 da \intddp \cr
&\ \  \Big[-\e_1\cdot (2\del_v-k_1) \e_2\cdot(2\del_v-k_1)\big]
e^{-T[p^2+a(k_1^2-2p\cdot k_1)]} e^{v\cdot p} \Big|_{v=0}  \cr
 = & \int_0^\infty T dT \int_0^1 da
 \Big[-\e_1\cdot (2\del_v-k_1) \e_2\cdot(2\del_v-k_1)
 (e^{ak_1\cdot v +v^2/4T}) \big] \Big|_{v=0} \cr
& \ \ \ \ \ \times\  e^{-Tk_1^2(a-a^2)} \intddpp e^{-Tp'^2}\ . \cr}
  \eqn\amassage $$
Carrying out the derivatives and the integral over momentum, and
adding to this expression the contribution of \diagramb, we are
left with
$$ \eqalign{
\Pi = &{-(g\mep)^2\over\fpe} \Tr(T^aT^b)\int_0^\infty \dToTe{1}  \cr
&\Big\{\Big[\int_0^1 da \Big(
-{2\over T}\e_1\cdot\e_2 - (1-2a)^2 \e_1\cdot k_1 \e_2\cdot k_1\Big)
e^{-Tk_1^2(a-a^2)}\Big]  \cr
&  \ \ \ \ \ \ \ \ \ \ \ \ \ \ \ +
{2\over T}\e_1\cdot\e_2\Big\} \ , \cr} \eqn\vacpoold $$
where $\e = 4-D$.

Alternatively we may write down the
result of \EAasBK\ for $N=2$:
$$ \eqalign {
 \Gamma_2(k_1,k_2) =
{(ig\mep)^2\over \fpe}& \linv \Tr(T^a T^b)
\loopTiNte{1}   \int_0^{1}  du\   e^{k_1\cdot k_2 G_B(1-u)} \cr
 \Big[&k_2\cdot \e_1
     k_1\cdot \e_2 [\Gdt_B(1-u)]^2
  +\ \e_1\cdot \e_2\  \Gdd_B(1-u)\big)\Big]\ .
\cr}
\eqn\vacpoBK $$
Define $a=1-u$, plug in the functions
in \GBdefn, and the result appears:
$$ \eqalign { \Pi = {-(g\mep)^2\over\fpe} &\Tr(T^aT^b)
\int_0^\infty \dToTe{1} \int_0^1 da\
 e^{Tk_1\cdot k_2 (a-a^2)}    \cr
&\Big[{2\over T}(\delta(a)-1)\e_1\cdot\e_2 +
(1-2a)^2 \e_1\cdot k_2 \e_2\cdot k_1
\Big] \ .  \cr}  \eqn\vacponew $$
Note that, as advertised, the diagram involving a four-point
vertex (eq.~\diagramb) is found by evaluating the delta function
in \vacponew; since $\Tr(T^aT^b)=\Tr(T^bT^a)$ this trace receives
the full contribution of the delta function.
This example also makes clear that,
as explained by \BnK\refmark\BKrules,
the differences $u_i-u_j$ are directly related to the usual
Feynman parameters.

\chapter{The Effective Action of a Spinor Particle
in a Background Field}

The case of a spinning particle is a simple generalization
of the particle theory used in section 2.
The one-loop action of a  Dirac
spinor with a vector-like coupling to a background field is
$$ S = \int d^4x\ \bar{\chi} (i\Dslash-m)\chi
  \eqn\backspin $$
where $D_{\mu} = \del_{\mu} - igA_{\mu}$.
The one-loop
effective action as a function of $A_{\mu}$ is therefore
$$ \eqalign{\Gamma[A]
& = \log\ \Big[\det\ \big( i\Dslash-m\big)\Big]  \cr
& = \half \log\
\Big[\det \big( i\Dslash-m\big)\ \det \big(-i\Dslash-m\big)\Big]  \cr
& = \half \log\ \Big[\det\ \big(D^2{\bf 1}\
-\ {ig\over 4} F_{\mu\nu} [\gamma^\mu,\gamma^\nu] + m^2\big)\Big]\ .
\cr } \eqn\effactFG $$
where I use $\det(\Dslash)=\det(\g^5\Dslash\g^5)=\det(-\Dslash)$.
This expression for the effective action is also associated with the
second-order action for a Dirac spinor
$$ S = \int d^4x -{1\over m}\chi_L^{\dag} (\Dslash^2+m^2)\chi_R^{}
  \eqn\backscnd $$
where the $\half$ in \effactFG\ appears because $\chi_{L,R}^{}$ are
two-component Weyl spinors.  The relevance of these formulas
to the \BK\ formalism was noted
by Bern and  Dunbar.\refmark\BDmap\

Since the gamma matrices are anticommuting operators, it is
natural to introduce worldline fermions to represent them.
This technique has long been employed
to introduce spin\refmark{\BdVH,\Berez,\Casspin},
and even color\refmark\Cascolor,
into quantum mechanics.  There is nothing mysterious about this;
finite representations of compact groups
can be generated by a set of fermionic operators.

One may therefore implement a supersymmetric generalization
of the procedure outlined in eq.~\effactSP, introducing Grassmann
fields $\psi^\mu(\tau)$ as partners of the fields $x^\mu(\tau)$.
I will want the usual fermionic anticommutation relations
$$ \{\psi^\m,\psi^\n\} = g^\mn \ ,\eqn\anticomm $$
which imply that as operators the $\psi^\m$ fields are just
constants equal to $\sqrt{\half} \gamma_\m$, and I
take as the Hilbert space of the theory the four components
$\ket{\a}$
of the Dirac fermion, which are acted on in the usual
way by the $\psi$ fields:
$$ \psi^\m\ket{\a} = {1\over \sqrt{2}} \gamma^\m_{\a\b}\ket{\b} \ .
\eqn\psiacts $$

I will now evaluate \effactFG\ (in the massless case)
as in section 2, taking
the worldline fermions to have the usual antiperiodic
boundary conditions.
(One need consider  periodic boundary conditions
only for chiral fermions\refmark\ThankLD.)
Direct construction of the particle path integral leads to
$$ \eqalign{\Gamma[A] &=
\half \Tr \log\ \Big[D^2{\bf 1}\
-\ {ig\over 4} F_\mn [\gamma^\m,\gamma^\n]\Big]\ \cr
&= -\half \tauint \sum_{\a} \intdfourp \cr
&\ \ \ \bra{\a,p}\exp\big[-{\half\EB T} \{(p+gA)^2
+ igF_\mn \psi^\m\psi^\n\}\big]\ket{\a,p} \cr
 &= -\half\tauint \cnst \int \Dx\ \Dpsi\ \cr & \ \ \
 \linv\Tr\
\exp\big[-\taupint (\otuE\xdt^2\ + \half\psi\cdot\dot\psi -
igA_\m\xdt^\m
+ ig(\EB/2)\psi^\mu F_{\mu\nu}\psi^\nu) \big] \ . \cr
}
     \eqn\effactFGb $$
The abelian version of this action was
first presented by Brink, Di Vecchia, and Howe\refmark\BdVH;
the nonabelian case was discussed by several
authors.\refmark\NAbCase\

In this way, the effective action for a spinor is expressed as a
{\it supersymmetric} Wilson loop, in a free supersymmetric
theory.  The particle action is invariant under the transformation
$$ \d_\eta x^\m = -\EB\eta\psi^\m \ ; \ \d_\eta\psi^\m = \eta \xdt^\m
\ . \eqn\fermSUSY$$
This supersymmetry and the superfield formulation of this theory have
been addressed by many authors, for example in reference \BdVH;
I will not discuss it further in this work.

Now let us consider the effective action \effactFGb\
at order $g^N$.
For the moment I shall ignore the $[A_\m,A_\nu]$ term in
$F_{\mu\nu}$; I will return to it at the end of this section.
 Expanding for the moment
 only the terms
with a single power of the gauge field to order $N$, and inserting
the momentum eigenstates of eq.~\Astate, one finds
$$ \eqalign {\Gamma^0[A] = -\half&{(ig)^N\over N!}
\tauint\ \cnst \int \Dx\ \Dpsi
 \exp[-\taupint (\otuE\xdt^2 + \half\psi\cdot\dot\psi)]\ \cr
 \linv&\Tr \prod_{i=1}^{N}
  \int_0^{\ T} dt_i\ \Big\{A_\m[x(t_i)]\cdot\xdt^\m(t_i)\ - \EB
  \psi^\m(t_i)\del_\m A_\n[x(t_i)]\cdot\psi^\n(t_i) \Big\}  \cr
 = -\half&(ig)^N \tauint\ \cnst \int \Dx\ \Dpsi
 \exp[-\taupint (\otuE\xdt^2 + \half\psi\cdot\dot\psi)]\ \cr
 \linv&\Tr \prod_{i=1}^{N}
  \int_0^{\ T} dt_i\
  T^{a_i}\ \big[\e_i\cdot\del_i x(t_i)\ + i\EB
  \e_i\cdot\psi(t_i)\ k_i\cdot\psi(t_i) \big]
  e^{ik_i\cdot x(t_i)}              \cr    }
\eqn\FEAorderN $$
(I write $\Gamma^0$ to remind the reader that I have left out
the commutator term in $F_\mn$.)
Here string theorists will find the vertex operators for vector
fields used in the superstring.

Again we can put the
polarization vectors in the exponentials; using Grassmann
variables $\ta$ and $\tab$, we may write
$$ \eqalign {\V \equiv\ & igT^{a}\ \big[\e\cdot\xdt + i\EB
  \e\cdot\psi\ k\cdot\psi \big]
  e^{ik\cdot x} \cr =\ &
  igT^{a}    \int d\ta d\tab
  \exp\big[\tab\ta\e\cdot\xdt +
  \ta\sqrt{\EB}\e\cdot\psi + i\tab\sqrt{\EB} k\cdot\psi +
  ik\cdot x \big] \ . \cr }  \eqn\Grassvtx $$
This leads to sources for $x^\m$
$$ J^\mu(\tau) = \sum_1^N\ \delta(\tau-t_i)
(\tab_i\ta_i\e_i^\mu \del_{t_i} + ik_i^\mu) \ .
\eqn\grasource $$
and $\psi^\m$
$$ \eta^\m(\tau,\ta,\tab) = \sum_1^N
\delta(\tau-t_i) \sqrt{\EB}(\ta_i\e_i + i\tab_i k_i)
\ .  \eqn  \Fsource  $$

The result of carrying out the $x$ and
$\psi$ integrals (in the gauge $\EB=2$) is
$$ \eqalign {  \Gamma^0_N(k_1,\dots,k_N)
= \Cb &\linv\Tr(T^{a_N}\dots T^{a_1})
\loopTiNt{3-N}
\Big(\prod_{i=1}^{N} \int_0^{\ u_{i+1}} du_i\ \Big)\cr
 \exp\Big(\sum_{i<j=1}^N& k_i\cdot k_j G_B^{ji} \Big)
\bigg\{\Big(\prod_{i=1}^{N}  \int  d\ta_i d\tab_i \Big) \cr
&\exp\Big(\sum_{i<j=1}^N \big(-i\
(\tab_j\ta_j k_i\cdot \e_j -  \tab_i\ta_i k_j\cdot \e_i)\Gbd{ji}    \cr
& \ \ \ \ \ \ \ \ +  \tab_i\ta_i\tab_j\ta_j  \e_i\cdot \e_j \Gbdd{ji}
\big)\Big)\cr
& \exp \Big(\sum_{i<j=1}^N\big[-\tab_i\tab_j k_i\cdot k_j
 \ + i \tab_i\ta_j k_i\cdot \e_j \cr
& \ \ \ \ \ \ \ \ \
 \ + i \ta_i\tab_j \e_i\cdot k_j
+\ \ta_i\ta_j\e_i\cdot \e_j\big]  G_F^{ji}\Big) \
  \bigg\}\  ,
\cr}
\eqn\EAwGBFs$$
plus terms involving all other proper-time/color orderings.
The overall factor of four comes from
$$ \int\Dpsi\ e^{-\taupint\half\psi\cdot\dot\psi} =
\Tr_{\psi}{\bf 1} = \sum_{\a=1}^4 \VEV{\a|\a} \ . \eqn\fourfctr $$

The generating kinematic factor (in braces) has a bosonic part
identical to \gkfdefn, as well as terms that contain
the one-loop Green functions
$G_F$ ($\Gf{ji}=-\Gf{ij}\equiv G_F(t_j-t_i)$)
of the fermionic $\psi$ fields.
In addition to implementing the constraint that
 every polarization vector appears
exactly once, the Grassmann integrations over $\ta$ and $\tab$
ensure that in any term of the generating kinematic factor
in which $\e_i^\m \Gf{ij}$ appears,
$k_i^\n \Gf{ik}$ must also appear.
This implies that the $G_F$ functions always
occur in closed chains of the form
$$ \prod_{k=1}^d \Gf{i_{k+1},i_k}\  ;
 \ (i_{d+1}\equiv i_1) \ .
\eqn\GFchain $$
(As the $G_F$'s are antisymmetric in their arguments,
a term like $\Gf{12}\Gf{13}\Gf{23}$ is not ruled out; on the contrary,
it is equal to $-\Gf{12}\Gf{23}\Gf{31}$ which is of the form \GFchain.)

The bosonic part of the action in \effactFGb\ is the
same as in section 2, so the $G_B$ functions are again given
by eq.~\GBdefn.
The $G_F$ functions satisfy
$$ \half\del_t G_F(t,t') =  \delta(t-t') .
\eqn\GFFdd $$
Since the fermions also satisfy antiperiodic boundary conditions
$$ \psi(t\rarr T) = -\psi(t\rarr 0)\ , \eqn\antifbc $$
we take the  antiperiodic solution of
eq.~\GFFdd:
$$ \ G_F(t,t') = \sign(t-t') = \sign(u-u').    \eqn\GFFloop $$
This function is double-valued, since it changes sign only at
$t=t'$:
$$ \ G_F(t,t') = - G_F(t+T,t')       .    \eqn\dblvalu $$

If the theory is abelian, then the single expression \EAwGBFs\
contains the entire one-loop effective action (which is also
the full photon one-loop S-matrix.)  However, if
we are working in a non-abelian gauge theory, then
in addition to the expression given in \EAwGBFs\ for the effective
action we must include terms involving the quadratic
term in $F_\mn$,
$$ -\half g^2\EB\psi^\m [A_\m,A_\n] \psi^\n \ ,   \eqn\AAterm $$
which generates two-gluon vertex operators
of the form
$$ \O_{i,j} =
-g^2\EB(T^{a_j}T^{a_i})\ \e_j\cdot\psi\ \e_i\cdot\psi\
e^{i(k_i+k_j)\cdot x}\ .
\eqn\AAvtxop $$
In the second-order formalism for spinors in gauge fields,
the usual three-point vertex is replaced by a new three-point
vertex along with a two-gluon/two-spinor vertex,
similar to the vertices of
scalars in gauge fields.  This can be inferred from eq.~\backscnd.
As in the previous section,
a part of the four-point vertex
is associated with the delta function in $\Gdd_B$,
but because of the particle's spin and the non-abelian nature of
the background field
this vertex contains a
new piece generated by the operator $\O_{i,j}$.

The contribution of this operator can be evaluated through a
process known as ``pinching'', which is related to the \BK\
rules for trees attached to loops.
In this process gluons $i$ and $j$
are brought to the same point on the loop (``pinched''), and a
subsidiary ``pinched kinematic factor'', containing the contribution
of $\O_{i,j}$, is extracted from
the generating kinematic factor in a systematic way.
The reader may wish to review the \BK\ rules\refmark\BKPascos,
which serve as motivation
for the following unusual manipulation of \AAvtxop:
$$ \eqalign {\O_{i,j} = & \
-g^2(T^{a_j}T^{a_i})\int\ d\ta_i d\tab_i d\ta_j d\tab_j \cr
& \ \ \ \ \ \ \ \ \ (-\tab_i\tab_j)
  \exp\big[
  \ta_i\sqrt{\EB}\e_i\cdot\psi(t_i)
  + \ta_j\sqrt{\EB} \e_j\cdot\psi(t_j) +
  i(k_i+k_j)\cdot x\big]\Big|_{t_i=t_j}\cr
= & \
(ig)^2(T^{a_j}T^{a_i})\int\ d\ta_i d\tab_i d\ta_j d\tab_j \cr
& \ \ \ \ \
 {\exp\big[
  \ta_i\sqrt{\EB}\e_i\cdot\psi + \ta_j\sqrt{\EB} \e_j\cdot\psi +
  i(k_i+k_j)\cdot x
-\tab_i\tab_j k_i\cdot k_j \Gf{ji}\big] \over
 k_i\cdot k_j \Gf{ji}}   \Bigg|_{t_i=t_j}              \cr}
\eqn\AArewrt $$
Insertion of this operator into  \FEAorderN\ to replace two operators
of the type \Grassvtx\ gives the pinched kinematic factor.
Comparison with \EAwGBFs\ shows that the {\it pinched} factor consists
of all the terms in the {\it generating} kinematic factor which
contain $k_i\cdot k_j \Gf{ji}$,
with the replacement
$$k_i\cdot k_j \Gf{ji} \rarr
\cases{+1,& if $t_j>t_i$; \cr -1,& if $t_j<t_i$ ,\cr}   \eqn\GFrplc $$
and with $t_i$ set equal to $t_j$.
Notice that if a term contains $\e_i\cdot \e_j \Gf{ji}$ as well,
it vanishes since
$\Gf{ji}(0)\equiv 0$
by antisymmetry.

In order to keep track of the different pinch contributions,
it is useful to write down a simple mnemonic rule based on
\BK\ diagrams.  While this could be done in many ways,
the particular choice presented here will eventually permit
a smoother transition from effective actions to
scattering amplitudes.

Draw all (planar) $\phi^3$ graphs with one loop,
$N$ external legs and any number $N_T\leq N/2$ of
trees with {\it one} vertex.
Consider a particular graph and a particular color(path)-ordering;
label the external legs clockwise from $1$ to $N$
following the path-ordering.
Now examine the
generating kinematic factor of \EAwGBFs\ term by term.
Two external gluons flow into each tree vertex;
let $j$ be the gluon lying most clockwise, and call
the other gluon $i$.
If a given term does not
contain a factor $k_i\cdot k_j \Gf{ji}$ for
{\it each} tree vertex in the graph,
then it vanishes.  Even then, it must contain exactly
{\it one} $\Gf{ji}$ at each vertex; otherwise it vanishes.  If it
survives, then replace each $k_i\cdot k_j\Gf{ji}$ by $+1$,
replace $t_i \rarr t_j$ in all Green functions, and eliminate the
$t_i$ integral.

As an application of the formalism of this chapter,
let us consider the contribution of a Dirac spinor
to the gluon vacuum
polarization.  In the usual first-order formalism of Dirac,
the single diagram has the form
$$ g^2 \Tr(T^aT^b) \intddp
 {-\Tr[\eslash_1(\pslash-\kslash_1)\eslash_2\pslash] \over p^2(p-k_1)^2}
  \eqn\diagramc $$
Usually this diagram is evaluated by writing
$$ \Tr[\eslash_1(\pslash-\kslash_1)\eslash_2\pslash] =
 4[ \e_1\cdot (p-k_1)\e_2\cdot p
  + \e_1\cdot p\ \e_2\cdot (p-k_1)
 - p\cdot (p-k_1) \e_1\cdot \e_2]\ ,
     \eqn\GamtrcA $$
after which the momentum integral is performed.  One may also use
$$ \eslash_i (\pslash-\kslash_i) = 2\e_i\cdot p\ -
\pslash\eslash_i
           - \eslash_i\kslash_i \eqn\Gamtrick $$
and write (after some algebra)
$$ \eqalign { 2\Tr[\eslash_1&(\pslash-\kslash_1)\eslash_2\pslash] \cr
& =
   \Tr[-\eslash_1\eslash_2](p^2+(p-k_1)^2)
+ \Tr[(2\e_1\cdot p - \eslash_1 \kslash_1)
    (2\e_2\cdot (p-k_1) - \eslash_2 \kslash_2)]  \cr
& = -4\ee{1}{2} \big(p^2+(p-k_1)^2\big) +
 4\e_1\cdot (2p-k_1)\ \e_2\cdot(2p-k_1) \cr &
\ \ \ \ \ - 4 \big(\ee{1}{2}\kk{1}{2} - \ek{1}{2}\ke{1}{2}\big) \cr}
    \eqn\GamtrcB $$
which puts the amplitude in a second-order form.
The first and second
term yield the contribution of \amassage\ times a factor of $-2$;
the last term is independent of the loop momentum.
The result is
$$ \eqalign{
\Pi = &\ 2{(g\mep)^2\over\fpe} \Tr(T^aT^b)\int_0^\infty \dToTe{1}  \cr
&\ \ \ \Big\{\Big[\int_0^1 da \Big(
-{2\over T}\e_1\cdot\e_2 - (1-2a)^2 \e_1\cdot k_1 \e_2\cdot k_1\cr
& \ \ \ \ \
 + \big(\ee{1}{2}\kk{1}{2} - \ek{1}{2}\ke{1}{2}\big)
\Big)
e^{-Tk_1^2(a-a^2)}\Big]  \cr
&  \ \ \ \ \ \ \ \ \ \ \ \ \ \ \ +
{2\over T}\e_1\cdot\e_2\Big\} \ , \cr} \eqn\VPFold $$
where $\e = 4-D$.

By contrast, evaluation of \EAwGBFs\ at order $g^2$ immediately
yields
$$ \eqalign {
 \Gamma_2(k_1,k_2) =
-2{(ig\mep)^2\over \fpe}& \linv \Tr(T^a T^b)
\loopTiNte{1}   \int_0^{1}  du\   e^{k_1\cdot k_2 G_B(1-u)} \cr
 \Big[& \e_1\cdot k_2      \e_2\cdot k_1 [\Gdt_B(1-u)]^2
  +\ \e_1\cdot \e_2\  \Gdd_B(1-u)\big) \cr
 &+ \big(\ee{1}{2}\kk{1}{2} - \ek{1}{2}\ek{2}{1}\big)[G_F(1-u)]^2
  \Big]\ ,
\cr}
\eqn\VPFBK $$
which is identical to \VPFold.
There are no pinches to perform, since the integrand contains
no terms with a single power of $\Gf{12}$.

\chapter{The Effective Action of a Vector Particle
in a Background Field}

Now let us consider the case of a massless spin-one particle.
There are many ways to proceed, and among them are several
directly inspired by the methods of string theory.
In a model inherited from
the bosonic string, one would  introduce
a single oscillator mode with a vector index, whose sole purpose
would be to excite an unphysical scalar
``vacuum'' (which would eventually be removed by hand)
to a vector boson state.  One could then imagine projecting out
all higher spin states, either by hand or by tricks ranging from
adding large masses (as in the string) or by adding complex phases
to the oscillators (along the lines of string orbifold constructions).
Another possibility is to use a supersymmetric construction; as in
the superstring, a fermionic oscillator with a vector index
can be used to excite a
``vacuum'' (which one projects away) to a state with vector indices.
Extra states can again be projected out in a number of ways.
I will use this latter construction, following closely both
the usual superstring methodology\refmark\GSW\
 and the work of Brink, Di Vecchia
and Howe.\refmark\BdVH\

The action of a Yang-Mills particle $Q_\m$, expressed in Feynman gauge,
in a classical background $A^\m$ is
well-known to be
$$ \eqalign {
 S = \int d^4x\ \{Q^{a\m} [(D^2)^{ab}g_{\m\n}& - g
(F^c_{\r\s}J^{\r\s})_{\m\n}f^{cab}] Q^{b\n}
 + \bar{\omega} (D^2)^{ab} \omega \cr
&+ {\rm order}(Q^3,Q^4,\bar{\omega}Q\omega,etc.)\}\ , \cr }
  \eqn\backyang $$
where $D_{\mu} = \del_{\mu} - igA_{\mu}$ and $gF_\mn = i[D_\m,D_\n]$
are functions {\it only} of the background field, $\omega$ is the
ghost of
background field Feynman gauge, and $J_\mn$ is the spin-one
(hermitean) generator of Lorentz transformations:
$$ (J_\mn)^{\r\s} =  i (\d^\r_\m \d^\s_\n - \d^\r_\n \d^\s_\m) \ .
\eqn\defnJmn $$
(Feynman gauge for $Q_\m$ is appropriate
in that the propagator is
$\square^{-1}$, as we had for scalars and spinors;
background field gauge is essential since the result must be
gauge invariant with respect to the classical field $A_\m$.
The appearance of background field gauge in this context
and the following expression for the effective action
were discussed in the work of Bern and Dunbar.\refmark\BDmap\
A useful introduction to background field gauge is
given in reference \Abbott.)
The one-loop effective action is found from the part of
\backyang\ which is quadratic in the quantum fields:
$$ \eqalign{\Gamma[A] = &-\half\log\ \Big[\det\ (D^2\
-\ g  F^\mn J_\mn) \Big]\ + \log\Big[\det\ (D^2)\Big]\ . \cr}
\eqn\EAvector $$
Again the structure of the effective action suggests the use of
Grassmann variables, and turning to Brink, Di Vecchia,
and Howe\refmark\BdVH,
we find that they have discussed the relevant theory.

Let us consider a particle with coordinates $(x^\m,\psp^\m,\psm^\m)$.
We will find it useful to consider also the real field
$\psi^\m=(\psm^\m+\psp^\m)$.  The worldline fermions satisfy
$$ \eqalign {
\{\psp^\m,\psm^\n\} &= g^\mn = \half\{\psi^\m,\psi^\n\} \ ; \cr
\{\psp^\m,\psp^\n\} &= \{\psm^\m,\psm^\n\} = 0 \ . \cr}
\eqn\vanticomm  $$
 If we define a vacuum $\ket{0} $ as the state
such  that $\psm^\m\ket{0}=0$ for all $\m$, then the full set of
sixteen states (for a given momentum) is
$$ \ket{0} ;\ \psp^\m\ket{0} ;\    [\psp^\m, \psp^\n]\ket{0} ; \
\e_{\m\n\r\s}\psp^\n\psp^\r\psp^\s\ket{0} ; \
\e_{\m\n\r\s}  \psp^\m \psp^\n \psp^\r \psp^\s \ket{0} \ .
\eqn\allstate  $$
These are antisymmetric tensors; in four-dimensions
 the (0,1,2,3,4)-index
 antisymmetric tensors have (1,4,6,4,1) components
of which only
(1,2,1,0,0) are physical degrees of freedom. This model therefore
describes a scalar, a vector boson, and a pseudoscalar.  However,
if we can implement a projection
onto states with odd fermion number, then the
truncated Hilbert space
$$ \psp^\m\ket{0} \ \ {\rm and}\ \
\e_{\m\n\r\s}\psp^\n\psp^\r\psp^\s\ket{0} \  .
\eqn\thestates $$
will contain only the spin-one states as physical modes.

In a complete analysis of this truncated model, one must study
the superreparametrization ghosts in order to derive
 the \BK\ rules; however, I have chosen to skirt
the issue of ghosts in this article.
 For the present paper it will be
sufficient to use a trick borrowed from string theory,
in which the gluon ghosts of field theory are accounted for
by hand, and in which the three-index tensor is
given a mass which is sent to infinity at the end of
the calculation.

Derivation of the superparticle Lagrangian is straightforward
when one observes the following:
$$ \bra{\r}{i\over 2}[\psi_\m,\psi_\n]\ket{\s} = {i\over 2}
   \VEV{\psm^\r[\psi_\m,\psi_\n]\psp^\s} = (J_\mn)^{\r\s} .
\eqn\Jmnfound $$
Remembering that we will eventually do away with the spurious states,
 let us extend the theory to the full set of
sixteen states in \allstate.  As in \effactFGb,
we are led to the particle Lagrangian
$$ \eqalign  { L = \otuE\xdt^2\
 + \psp\cdot\dot\psm - igA_\m\xdt^\m
+ ig{\EB\over 2}\psi^\mu F_{\mu\nu}\psi^\nu\ . \cr
 }   \eqn\vectLgr $$
However, in order to carry out the trick described above
we will want to make the three-index tensor heavy.
We must therefore break the degeneracy of the sixteen states
by adding a harmonic oscillator
potential:
$$ L \rarr L  - C(\psp\cdot\psm-1) \ .        \eqn\vectwSHO $$
For positive $C$ the $\psi$'s form a
fermionic harmonic oscillator whose states are spaced by
$\Delta m^2=C$ and whose vacuum is a tachyon with $m^2=-C$.
Fortunately this tachyon is unphysical;
it will be removed from the theory by truncation as discussed above,
and so causes no difficulties.
  All other states except the vector boson will vanish
as a result of the truncation or because
their masses will be taken to infinity.
(This construction is taken directly from the
superstring.\refmark\GSW)

One can proceed straightforwardly with the computation of
the effective action in direct analogy to the spinor and scalar
cases.  The field theory ghosts in background
field gauge contribute a factor of $\log \det D^2$; as noted
by Bern and Dunbar\refmark\BDmap, and as expected from string
theory\refmark\GSW, this is
exactly the negative of the effective action of a complex scalar
in the adjoint representation (see eq.~\effactSG):
$$
\Gamma[A]_{\rm ghosts} = - \tauint\ \cnst \int \Dx\  \Tr
\exp\Big[-\taupint (\otuE\xdt^2\
-\ igA\cdot\xdt)\Big]       \ .
\eqn\EAghost $$
The gauge boson contribution may  be calculated by projecting out
the even fermion states in the theory and by letting $C\rarr\infty$.
The projection, which is the GSO projection well-known
from string theory\refmark\GSO, is implemented by the operator
$$P_{GSO} = \half\big[1-(-1)^F\big]\ , \eqn\PGSO $$
where $F=(\psp)^\m\cdot (\psm)_\m$ is the fermion
number of a state.  Clearly only the states of \thestates\ survive.
It is well-known\refmark\GSW\ that the operator $(-1)^F$ is
 implemented in the path-integral by choosing periodic boundary
conditions for fermions:
$$ \psi(t\rarr T) = \psi(t\rarr 0) \ . \eqn\perifbc $$
We may therefore write
$$ \eqalign{\Gamma[A] &=
              -\half \Tr \log\ \Big[D^2{\bf 1}\
-\ g F^\mn J_\mn \Big]\ \cr
&= \half\lim_{C\rarr\infty}
 \tauint \sum_{s_0,s_1,s_2,s_3=0}^1 \intdfourp \cr
& \ \ \ \bra{s_\r,p}P_{GSO}
\exp\big[-{\half\EB T} \{(p+gA)^2 \cr
& \ \ \ \ \ \ \ \ \ \ \ \ \ \ \ \ \ \ \ \ \ \ \ \ \ \ \ \ \ \ \ \ \
- C(\psp\cdot\psm-1)
+ igF_\mn \psi^\m \psi^\n\}\big]\ket{s_\r,p}   \cr
 &= \half\lim_{C\rarr\infty}
 \tauint\ \cnst \int \Dx\
 \half\Big[\int_{(\half)}\Dpsi\
         -  \int_{(0)} \Dpsi\ \Big]
\cr & \ \ \
 \linv\Tr\
\exp\big[-\taupint\ (\otuE\xdt^2\ + \psp\cdot\dot\psm
    - {\EB\over 2}C(\psp\cdot\psm-1) \cr
&\ \ \ \ \ \ \ \ \ \ \ \ \ \ \ \ \ \ \ \ \ \ \ \ \ \ \ \ \ \ \ \ \ \ \ \
-igA_\m\xdt^\m
+ ig{\EB\over 2}\psi^\mu F_{\mu\nu}\psi^\nu ) \big] \ , \cr }
     \eqn\EAmodel $$
where the subscripts ($\half$) and ($0$) indicate antiperiodic and
periodic boundary conditions on the worldline fermions.

Proceeding as in the previous section (eqs.~\effactFGb--\EAwGBFs),
we  find (in the gauge $\EB=2$)
$$ \eqalign {
 \Gamma^0_N(k_1,\dots,k_N)
= \Cc &\linv\Tr(T^{a_N}\dots T^{a_1})
\loopTiNt{3-N}
\Big(\prod_{i=1}^{N} \int_0^{\ u_{i+1}} du_i\ \Big)\cr
 \exp\Big(\sum_{i<j=1}^N &k_i\cdot k_j G_B^{ji} \Big)
\bigg\{\Big(\prod_{i=1}^{N}  \int  d\ta_i d\tab_i \Big) \cr
&\exp\Big(\sum_{i<j=1}^N \big(-i\
(\tab_j\ta_j k_i\cdot \e_j -  \tab_i\ta_i k_j\cdot \e_i)\Gbd{ji}    \cr
& \ \ \ \ \ \ \ \ +  \tab_i\ta_i\tab_j\ta_j  \e_i\cdot \e_j \Gbdd{ji}
\big)\Big)\cr
\sum_{p=0}^1 \ (-)^{p+1} & {Z_{{p\over 2}}\over 2}
 \exp \Big(2 \sum_{i<j=1}^N\big[-\tab_i\tab_j  k_i\cdot k_j
 \ + i \tab_i\ta_j k_i\cdot \e_j \cr
& \ \ \ \ \ \ \ \ \ \ \ \ \ \
 \ + i \ta_i\tab_j \e_i\cdot k_j
+\ \ta_i\ta_j\e_i\cdot \e_j\big]  G_F^{({p\over2})ji}\Big) \
  \bigg\}\  ;
\cr}
\eqn\VEAwGBFs$$
again the symbols ($\half$) and ($0$) indicate antiperiodic and
periodic fermions.  Notice the factor of two relative to \EAwGBFs\
in the exponential of the fermionic Green functions.
The $Z$ factors are given (in Minkowski spacetime) by
$$ \eqalign{ \left\{ \matrix{Z_{(\half)}\cr Z_{(0)}\cr} \right\}
=& \int_{\psi(T)=(\mp)\psi(0)}
\Dpsi\ e^{-\taupint [\psp\cdot\dot\psm - C(\psp\cdot\psm-1)]} =
\Tr_\psi (\pm 1)^F e^{-H[\psi]T}  \cr
  = & e^{-CT} \big(\sum_{s=0}^1
  \bra{s}(\pm e^{CT})^s\ket{s} \big)^4
  = 16e^{CT} \left\{ \matrix{\ch^4\cr \sh^4 \cr} \right\}
  (-CT/2)     \cr
  = & e^{-CT} \pm 4 + 6e^{CT} + \dots
\cr } \eqn\Zfactor $$
When continued to Euclidean spacetime, the
arguments of  the
exponentials change sign; cancellations remove
all growing exponentials, as I will explain below.

The bosonic green
functions are identical to those used for the scalar and spinor
particle (eq.~\GBdefn), since the free bosonic action
$$L_B = \otuE \xdt^2 \eqn\LgrB$$
is independent of the particle's
spin. The free fermionic action is
$$ L_F = \psp\cdot\dot\psm -  C\psp\cdot\psm \ ; \eqn\LgrF $$
however, this leads in Minkowski
spacetime to Green functions which blow up as $C\rarr\infty$.  It is
therefore necessary to analytically continue to Euclidean
spacetime to study this limit.

Moving to Euclidean spacetime, and being careful to define
the number operator properly, we have
$$ L_F^{Eucl} = \psp^\m g_\mn(\del_t +  C)\psm^\n \ , \eqn\LgrFEucl $$
Let us first compute the Green functions on the line. Define
$$\GFpm(t,t') = \VEV{\psp(t)\psm(t')} \ ; \eqn\defngfpm  $$
this function satisfies
$$ (\del_t + \ta(t-t')C)\GFpm(t,t') = \delta(t-t') \ . \eqn\GFpmeqn $$
where $\theta(t)$ is a step function which is zero for negative $t$.
This equation implies
$$\GFpm(t,t')  = \ta(t-t') \exp(-C|t-t'|)\ .  \eqn\GFpmline  $$
Similarly
$$\GFmp(t,t')  = -\ta(t'-t) \exp(-C|t-t'|)\ .  \eqn\GFmpline  $$
Since $\GFpp$ and $\GFmm$ both vanish,
$$ G_F(t,t') = \VEV{\psi(t)\psi(t')}
       =  \sign(t-t') \exp(-C|t-t'|)\ .  \eqn\GFlineV $$
On the circle of circumference $T$,
we will need to find functions,
one periodic
($\GFp$), another antiperiodic ($\GFa$) in $t\rarr t+T$,
which reduce to eq.~\GFlineV\
in the limit that $T\rarr\infty$.  An analysis analogous to the
above yields
$$ \eqalign{
\GFa(t-t') &=  2\ \sign(t-t') e^{-\half CT}\
\ch\big[C(\half T - |t-t'|)\big]\ ;\cr
\GFp(t-t') &=  2\ \sign(t-t') e^{-\half CT}\
\sh\big[C(\half T - |t-t'|)\big]
\ .\cr } \eqn\GFcircV $$
Again these are precisely the functions found by Bern and Kosower
in the derivation of their field theory rules\refmark\BKrules.

The next task is to discard the three-index tensor by sending $C$
to infinity.
We must carefully analyze the effective action \VEAwGBFs\
to see what terms remain in this limit.
The following discussion is almost identical to that of
\BnK\refmark\BKrules; I repeat it here for the sake of completeness.

It is necessary to study
separately terms with and without $G_F$ chains.
For terms in \VEAwGBFs\ that contain no $G_F$'s, the only
dependence on $C$ is given in the prefactors $Z_{{p\over2}}$, which
in Euclidean spacetime take the form
$$  \left\{ \matrix{Z_{(\half)}\cr Z_{(0)}\cr} \right\}
  = 16e^{CT} \left\{ \matrix{\ch^4\cr \sh^4 \cr} \right\} (CT/2)
  =  e^{CT} \pm 4 + 6e^{-CT} + \dots
 \eqn\EuclZs $$
  The first
term, associated with the propagation of the tachyon,
 blows up as $C\rarr\infty$; fortunately it cancels
in the expression
$$\half  \big(Z_{(\half)}- Z_{(0)}\big) = 4 + \O(e^{-2CT})\  ,
\eqn\tachygon $$
leaving us with an overall factor of 4.
This factor stems from the
sum over the four states $\psp^\m\ket{0}$ which can propagate around
the loop.  These purely bosonic terms are partially
cancelled by the contribution of the ghosts (eq.~\EAghost);
the removal of the
timelike and longitudinal modes of the vector boson
reduces the number of states, and the overall factor, from 4 to 2.
(In the usual dimensional regularization schemes, this number
becomes $2-\half\e$; however it is natural in this formalism to use
dimensional reduction  or the variant of it developed
by \BnK\refmark{\BKrules,\BKPascos}, in which
the number of states is left at 2.)

Consider next the expansion in powers of $e^C$ of a chain product
of antiperiodic $\GFa$'s,
minus the same chain of periodic $\GFp$'s.  This is precisely
the sort of expression we obtain from \VEAwGBFs\ as a result
of the GSO projection.
{}From \GFcircV\ we find that
$$ \eqalign {\half\Big[\prod_{k=1}^d \GFa(t_{i_{k+1}},t_{i_k}&)
           - \prod_{k=1}^d \GFp(t_{i_{k+1}},t_{i_k}) \Big] \cr
= [\prod_1^d &\sign(t_{i_{k+1}}-t_{i_k})]
             e^{-CT}
\exp\big(-C\sum_{k=1}^d|t_{i_{k+1}}-t_{i_k}|\big) \times \cr
&\big[\sum_{n=1}^d
\exp\big(2C|t_{i_{n+1}}-t_{i_n}|\big) +
\O(e^{-CT})\big]  \ .\cr   }
\eqn\GFprod $$
(Here $i_{d+1}\equiv i_1$.)
            The leading term in \GFprod\ is of the form
$$
 [\prod_1^d \sign(t_{i_{k+1}}-t_{i_k})]              e^{-CT}
\sum_{n=1}^d \exp\big(-Cf(t_i;t_n)\big)
\eqn\leading$$
where
$$f(t_i;t_n) =
\sum_{k=1}^d|t_{i_{k+1}}-t_{i_k}|  -
 2|t_{i_{n+1}}-t_{i_n}| \geq 0
\  .     \eqn\fdefn $$
Unless $f(t_i;t_n) = 0$ for some $n$,
\leading\ will contribute too strong a
power of $e^{-C}$, and a term containing it
will vanish in the limit $C\rarr\infty$.

    Since the expressions above are
cyclic in $k$, one can rotate the $k$'s to make
    $t_{i_d}=t_{\max}\equiv\max[t_{i_k}]$;
let $t_{\min} \equiv \min[t_{i_k}]$. Then
$$
 2|t_{i_{n+1}}-t_{i_n}| \leq 2 (t_{\max}-t_{\min})
\leq\sum_{k=1}^d|t_{i_{k+1}}-t_{i_k}|  \  .
\eqn\manyts $$
For $f(t_i;t_n)=0$, both equalities must obtain.
Notice that the second equality can be satisfied only when
$$ t_{\max}=t_{i_d}>t_{i_{d-1}}> \cdots >t_{i_2}>t_{i_1} = t_{\min}
\eqn\ordereda $$
or
$$ t_{\max}=t_{i_d}>t_{i_1}>t_{i_2}>\cdots > t_{i_{d-1}} = t_{\min}
\ . \eqn\orderedb $$
 (I will call a chain
satisfying \ordereda\ or \orderedb\ a path-ordered chain
since the ordering is with respect to proper time.  I remind
the reader that the color trace is ordered in the same way.)
The first equality in \manyts\ can only hold when
$t_n=t_{\max}$  and
$t_{n+1}=t_{\min}$, or {\it vice versa}.
Thus the condition $f(t_i;t_n)=0$
 can only occur either when \ordereda\ holds and $n=d$,
or when \orderedb\ holds and $n=d-1$.
(In the case $d=2$, both
\ordereda\ and \orderedb\ hold.)
It follows that a path-ordered chain of $G_F$'s contributes
$$  [\prod_1^d \sign(t_{i_{k+1}}-t_{i_k})]  e^{-CT}
= e^{-CT}\cases{ & $-1$,  if \ordereda\ holds;
 \cr & $-(1)^{d-1}$,  if \orderedb\ holds;
 \cr & $-2$, if $d=2$. \cr }
\eqn\contribu    $$
Of course, as this derivation is essentially the same as
that of reference \BKrules,
the result \contribu\ agrees with that of \BnK.

  The exponential in \contribu\ cancels
the overall factor of $e^{CT}$ which was found in
eq.~\EuclZs, leaving only the numerical factor $-2$ or $\pm1$.
All other terms from such a chain, as well as those from
chains which are not path-ordered, have additional
decaying exponentials which vanish in the limit
$C\rarr\infty$.
Using the above argument twice, it is easy to see that a term
with more than one $G_F$ chain will always vanish
in the limit $C\rarr\infty$.
We therefore find that out of the expression
\VEAwGBFs, only terms with single
path-ordered chains of $G_F$'s of length $0$ to $N$
contribute, and then
are simply replaced by the factor $\pm1$ or $\pm2$.
At this point all dependence on $C$ has vanished and
we may return to Minkowski spacetime.

How should one interpret these rules? It is easiest to do so
from an operator standpoint.  Since we are
throwing away all states of \allstate\
except the spin-one tensor, we require that
the application of a $\psp$ operator, which moves us out of the
space of
spin-one states, be accompanied by the simultaneous application
of a $\psm$ operator in order
to bring us back to it.
This translates into a requirement that the Wick contractions
which generate the Green functions do
not overlap one another; hence the $G_F$'s must
be path-ordered.

We now have enough information to write down
a set of rules for the unpinched diagram, starting with the same
formula we had in the spinor case (eq.~\EAwGBFs).
To obtain the generating kinematic factor of the vector boson,
manipulate the kinematic factor of \EAwGBFs:
throw away all terms except
those with no $G_F$'s and those with a single $G_F$ chain, and
multiply terms without $G_F$'s by 2.
Next, replace the $G_F$ chains by
$$  [\prod_1^d \Gf{i_{k+1},i_k}]
\rarr \cases{ & $-2^d$,  if \ordereda\ holds;
 \cr & $-(-2)^d$,  if \orderedb\ holds;
 \cr & $-8$, if $d=2$; \cr & $0 \ $  otherwise. \cr}
\eqn\newcntrb    $$
where the powers of two account for the slight differences
between equations \VEAwGBFs\ and \EAwGBFs.  Finally,
substitute
the bosonic Green functions of \GBdefn, plug the result
back into \EAwGBFs, multiply by $-\fourth$ and evaluate the integral.

The non-Abelian part of $F_\mn$ contributes to amplitudes for
vectors just as it does for spinors.
The resulting pinch rules are almost as
described in the previous section, but
one must decide whether to
perform pinches before or after requiring that all
chains be path-ordered.  The relevant consideration is that
the pinch technique is just a trick
to generate the correct set of $G_F$'s;
one could drop the trick and calculate directly
the pinched kinematic factor by inserting
$\O_{i,j}$ (eq.~\AAvtxop)
into the path integral, just as is done in \FEAorderN\
with the usual $\V$'s (eq.~\Grassvtx).
Only {\it after} the whole set of $G_F$ chains in the pinched
kinematic factor is known
should one apply the analysis of eqs.~\GFprod--\contribu\
 to determine which chains survive in the limit $C\rarr\infty$.
Therefore, one should perform all pinches {\it before}
requiring that $G_F$ chains be path ordered; for example, the chain
$$ \Gf{12} \Gf{2,i+1} k_i\cdot k_{i+1} \Gf{i+1,i} \Gf{i1}
\eqn\badchain $$
for $t_N>t_{N-1}>\cdots>t_1$ will contribute to the diagram
in which gluons $i+1$ and $i$ are pinched, even though
in the evaluation of the unpinched \BK\ diagram
it is discarded.  (Notice that pinching cannot change the number
of $G_F$ chains in a given term, and so one may safely discard
from the original generating kinematic factor
any term with more than one such chain.)

Thus, the rule for pinched diagrams is the following:
Return to the generating kinematic factor
for the vector boson, and carry out the
pinches as explained in section 3.  Next, apply the
path-ordering requirement to $G_F$ chains, replacing them
with the factors in eq.~\newcntrb.  Finally, substitute the usual
functions for the $G_B$'s,  insert the kinematic factor into
\EAwGBFs, multiply by $-\fourth$
 and compute the integrals.

As an example, consider the pure $SU(N)$ Yang-Mills vacuum
polarization in background field gauge.  The reader may check
that if the algebra of Feynman diagrams is organized as explained by
 Bern and Dunbar\refmark\BDmap, it is straightforward to obtain
$$ \eqalign{
\Pi = &
{(g\mep)^2 f^{acd}f^{bdc}\over\fpe}\int_0^\infty \dToTe{1}  \cr
&\Big\{\Big[\int_0^1 da \Big(
-{2\over T}\e_1\cdot\e_2 - (1-2a)^2 \e_1\cdot k_1 \e_2\cdot k_1\cr
& \ \ \ \
 + 4 \big(\ee{1}{2}\kk{1}{2} - \ek{1}{2}\ke{1}{2}\big)
\Big)
e^{- Tk_1^2(a-a^2)}\Big]  \cr
&  \ \ \ \ \ \ \ \ \ \ \ \ \ \ \ \ \ \ +
{2\over T}\e_1\cdot\e_2\Big\} \ , \cr} \eqn\VPVold $$
where $\e = 4-D$. I have included the ghosts in this expression,
using dimensional reduction
in which the number of physical helicity
states is exactly 2.

According to the above rules for vector bosons,
this result can be extracted from the result of \VPFBK\ by
replacing $(\Gf{21})^2=-\Gf{21}\Gf{12}$ with $+8$,
multiplying the terms with
$(\Gbd{21})^2$ and $\Gbdd{21}$ by 2, and multiplying the entire
expression by $-\fourth$.  Indeed this gives
$$ \eqalign {
 \Pi =
 -{(g\mep)^2\over \fpe}& \linv \Tr(T^a T^b)
\loopTiNte{1}   \int_0^{1}  du\   e^{k_1\cdot k_2 G_B(1-u)} \cr
 \Big[& \e_1\cdot k_2      \e_2\cdot k_1 [\Gdt_B(1-u)]^2
  +\ \e_1\cdot \e_2\  \Gdd_B(1-u) \cr
 &+ 4\big(\ee{1}{2}\kk{1}{2} - \ek{1}{2}\ek{2}{1}\big)
  \Big]\ ,
\cr}
\eqn\VPVBK $$
which is identical to \VPVold\
(recall that $(T^a_{adj})^{cd} = -if^{acd}$.)
There are no pinches to perform; this is the complete result.

It is amusing to combine the results of \vacpoBK, \VPFBK\ and
\VPVBK.  Consider the gluon vacuum polarization
in a theory with $n_f$ Dirac fermions and $n_s$ complex scalars in
the adjoint representation:
$$ \eqalign {
 \Pi =
 -{(g\mep)^2\over 2\fpe}& \linv \Tr(T^a T^b)
\loopTiNte{1}   \int_0^{1}  du\   e^{k_1\cdot k_2 G_B(1-u)} \cr
\bigg\{(2-4n_f+2n_s)
 \Big[& \e_1\cdot k_2      \e_2\cdot k_1 [\Gdt_B(1-u)]^2
  +\ \e_1\cdot \e_2\  \Gdd_B(1-u) \Big]\cr
 &+ 4(2-n_f)\big(\ee{1}{2}\kk{1}{2} - \ek{1}{2}\ek{2}{1}\big)
  \bigg\}\ ,
\cr}
\eqn\VPsusy$$
(Since $\g^5$ does not play a role in vacuum polarizations,
the contribution of a chiral fermion to the above expression
is exactly half that of a Dirac fermion.)
Notice that the factor multiplying
the bosonic Green functions counts degrees of freedom, and
therefore cancels for all supermultiplets.  With appropriate
choices of matter supermultiplets in various representations,
it is possible to make the remainder of
\VPsusy\ vanish, leaving the theory
one-loop finite.  When all particles are
in the adjoint representation, complete cancellation occurs for
the case $n_f=2$ and $n_s=3$;
this is the famous
$N=4$ spacetime supersymmetric Yang-Mills theory,
which is known to be finite.\refmark\West\
Notice that this result requires no integrations; it
follows directly from the rules for obtaining the generating
kinematic factors from \EAwGBFs\ and from the overall normalizations.

\chapter{Integration by Parts and Manifest Gauge Invariance}
Bern and Kosower\refmark{\BKrules,\BKcolor} showed that there
are benefits associated with performing an integration-by-parts (IBP)
on all terms involving a $\Gdd_B$; when the $\Gdd_B$'s
are completely eliminated,
it is possible to derive a much simpler  set of rules
for scattering amplitudes.
As discussed by Bern and Dunbar\refmark\BDmap, this IBP causes
an interesting and intricate reshuffling of terms.
Essentially, the delta-functions which produce the four-point
vertices of field theory are removed by the IBP, allowing
a scattering amplitude to be expressed in terms of \BK\ graphs,
which have only $\phi^3$ vertices.
Each \BK\ graph is related to the ``unpinched diagram'' -- the
one with all gluons attached directly to the loop -- through the
systematic pinch prescription.

In the effective action, the reorganization from the
IBP is not much of
a simplification, as it leads to as many or more diagrams than
Feynman graphs.
Nonetheless it is worthwhile in many
cases: the additional diagrams are
easier to calculate
than usual Feynman graphs due to the systematic ``pinch'' rules,
and the number of types of  Feynman parameter integrals is reduced.
Furthermore, and perhaps most importantly,
it makes possible a direct analysis
of individual gauge invariant
 contributions to the effective action.
Still, the IBP is not essential for effective actions, and the
casual reader may safely skip this section at a first reading.

The reader intending to study this section
should be warned that the IBP,
while necessary for a complete picture of the possibilities
opened by the work of \BnK, represents the weakest link in the
present paper. A full understanding of the IBP
requires a clarification of the role of string duality,
which permits the reorganization which I will outline below.
In the absence of this clarification it is only
possible to present the IBP and the associated
pinch rules as
a trick, motivated by the \BK\ rules for scattering
amplitudes\refmark{\BKrules,\BKPascos} and the work
of Bern and Dunbar.\refmark\BDmap\
Specifically, these rules
match on to the \BK\ rules when the external
gluons are on-shell.
I will demonstrate the validity of this trick in a simple
case; however, while
I have checked that it works in more complicated cases,
I do not know a complete proof.
For this reason these effective-action pinch rules
appear completely {\it ad hoc} at the present time, and
the reader is urged to familiarize
herself with the \BK\ rules outlined in reference \BKPascos\
to help put the present section in context.

To illustrate the trick, I present the simplest case.  Consider
a term from the generating kinematic factor of \EAwGBFs\
of the form
$$ \ee{i}{j} \Gbdd{ij} \times F(\e_m,k_n) \ , \eqn\Ksimple $$
where $F$ contains neither $k_i$ nor $k_j$ and therefore has
no dependence on either $t_i$ or $t_j$.
The IBP of \Ksimple\ can be done with respect
to $t_i$, $t_j$, or $t_i-t_j$; different results will be
found in the different cases, the variations among them being
total derivatives.  For simplicity let us IBP with
respect to $t_i$; for a particular color ordering,
the initial expression from  \EAwGBFs\ is
$$ \tNint\cdots\tint{i}{i+1}\tint{i-1}{i}\cdots\tint{1}{2}
    \  \ee{i}{j}\ \Gdd_B(t_i-t_j) \ F \expkk
     \eqn\beforeIBP $$
which becomes
$$ \eqalign {
\tNint\cdots\tint{i}{i+1}&\tint{i-1}{i}\cdots\tint{1}{2}
    \ \ee{i}{j}\ \dot G_B (t_i-t_j) \ F \expkk \cr
\times  &\big[\delta(t_{i+1}-t_i) - \delta(t_i-t_{i-1})
               \ - \sum_{m\ne i} k_i\cdot k_m \dot G_B(t_i-t_m) \big]
\ . \cr }      \eqn\afterIBP $$
The last term now fits in neatly with the  terms in the generating
kinematic factor which lack $\Gdd_B$'s,
but the delta functions ---
the surface terms from the IBP --- are an annoyance.
(These delta functions contribute only to one color trace, so there
are no subtle factors of two associated with them.)
Essentially they are color commutators; they would cancel against
 surface terms from other proper-time orderings were the theory abelian,
but cannot do so here since different proper-time orderings have
independent color traces.
Fortunately these surface terms  bear
a simple relationship to the last term in \afterIBP.
Specifically, take the terms in the sum over $m$ with $m=i\pm1$:
$$ \eqalign {
- \tNint&\cdots\tint{i}{i+1}\tint{i-1}{i}\cdots\tint{1}{2} \cr
    &  \ee{i}{j} \dot G_B (t_i-t_j)
      \sum_{m=i\pm 1} k_i\cdot k_m \dot G_B(t_i-t_m)
    \ F \expkk    \ .
\cr }      \eqn\bfrpinch $$
Now, motivated by the pinch rules of section 3 and the work of
Bern and Dunbar\refmark\BDmap,
replace $k_i\cdot k_{i\pm1}\Gbd{i,i\pm1}$ with $\mp1$ and
set $t_{i}=t_{i\pm1}$; in this way the surface terms
are reproduced.

The case $j=i\pm1$ is special:
one of the surface terms contains $\Gbd{jj}\equiv 0$,
and so the pinch $t_i=t_j$
does {\it not} get a contribution from
the IBP.  This leads to a modification of the rule for ``pinching'':
the pinch of a term containing $(\Gbd{i,i\pm1})^2$ vanishes.  (Again
this matches with \BnK\refmark\BKPascos and with section 3.)
Recall that $\Gbdd{ij}$ contains a delta function,
which accounts for the Feynman graph in which a
four-point
vertex connects gluons $i$ and $j$; the missing surface term
is cancelled by the half of this
delta function that
contributes to the color trace under consideration.

In addition to terms like \Ksimple, the kinematic factor of
eq.~\EAwGBFs\ has terms in which $F(\e_m,k_n)$
contains $\Gdt_B$ functions dependent on $t_i$ and $t_j$, or in which
there are several $\Gdd_B$'s; these cases must be dealt with
in turn.
It appears that the resulting pinches are governed by simple rules,
which I will now present.  However, as mentioned above, no
proof exists for these rules; their main feature is their similarity
to the Bern-Kosower rules.

The first stage of the IBP reorganization involves the elimination
of all $\Gdd_B$'s in analogy to eqs.~\beforeIBP--\afterIBP.
Specifically, carry out the IBP of the generating kinematic factor,
{\it dropping all surface terms}, until no $\Gdd_B$'s remain.
(\BnK\ have proven that
this is always possible.\refmark\BKcolor)  The result
 is the ``improved generating
kinematic factor'', associated with the unpinched diagram.
Every term in this improved kinematic factor contains a
certain number of factors of $k_i\cdot k_j$, where $i$ and $j$ are
arbitrary.  The number of these factors cannot exceed
 $N/2$, since the maximum number of
$\Gbdd{ij}$'s and $k_i\cdot k_j\Gf{ij}$'s in any term in the
original generating kinematic factor is also $N/2$.
  Each pinch
absorbs one of these factors, as well as one of the integrals
over $t_i$, and so the maximum number of
pinches which must be performed simultaneously is $N/2$.

If the theory is abelian, then no further calculation is
necessary, as all surface terms do in fact vanish.
However, in a non-abelian theory, it is necessary to use the
following  pinch rules in order to
account for the IBP surface terms.
The procedure is
closely related to the \BK\
rules for scattering amplitudes; the reader is again urged
to review reference \BKPascos.

Draw all (planar) $\phi^3$ graphs with one loop,
$N$ external legs and any number $N_T$ of
trees, such that although each tree may have several vertices,
the total number of tree vertices $N_V$ is at
most $N/2$.
(Diagrams with trees may
seem out of place in the construction of a 1PI object like an
effective action, but the trees used here, unlike those
for scattering amplitudes, do not contribute the usual
propagator poles;
they serve only as a mnemonic
for ensuring all surface terms are accounted for.)
The gluons which flow into a tree before entering the loop are
said to be pinched; the number of these is $N_V+N_T$.
Consider a particular graph and a particular color(path)-ordering;
label the external legs clockwise from $1$ to $N$
following the path-ordering.
Each tree vertex, since it is a three-point vertex, is
characterized by one line pointing toward the loop and
two outward pointing lines $I$ and $J$, with
two sets of external legs $i_1,...,i_m$ and $j_1,...,j_n$
that flow into them.  Let $J$ be the line lying most clockwise.
Now examine the
improved generating kinematic factor term by term.
If a given term does not
contain a factor $k_i\cdot k_j \Gbd{ji}$ or $k_i\cdot k_j \Gf{ji}$
for {\it each} tree vertex, where $i$
belongs to the set of gluons flowing into
line $I$ and $j$ flows into $J$,
then it vanishes.  Even then, it must contain exactly
{\it one} $\Gbd{ji}$ or $\Gf{ji}$
at each vertex; otherwise it vanishes.  If it
survives, then
 replace $\Gbd{ji}$ or
$\Gf{ji}$ by $+1$, replace
$t_i\rarr t_j$ in all Green functions, and eliminate the
$t_i$ integral.
Finally, for every internal tree line (into which
flows momentum from gluons $r, r+1, \dots, s$), divide by
$$ \half\Big[(\sum_{q=r}^s k_q)^2 -
    \sum_{q=r}^s (k_q)^2\Big] , \eqn\funnypole $$
which becomes the expected intermediate-state pole only
when all external gluons are on-shell.  The effect of this
procedure is to produce contact terms; no actual poles
are ever generated.

It is useful to review the arguments of Bern and Kosower
for carrying out the IBP.\refmark{\BKrules,\BKPascos,\BKother}
After the IBP, the improved generating kinematic factor
is made up of only $\Gdt_B$'s and $G_F$'s; it has
no singularities and contains no dependence on $T$.   This
simpler form leads to
fewer separate integrations, and also
allowed \BnK\ to construct a
formalism in which one needs only $\phi^3$ graphs
to compute scattering amplitudes.
In addition, since the
kinematic factor is independent of $T$,
the overall power of $T$ is
given by the number of $t_i$ integrations;
a diagram with $N$ gluons and
$k$ pinches has an integral $\int dT/T^{3-N+k}$.
As a consequence,
the ultraviolet infinities of gauge theory
appear only in terms with $N-2$ pinches,
since $\int dT/T$ is the only possible source of ultraviolet
divergences.
Indeed one may interpret this
reorganized amplitude using gauge invariant
structures.   I will illustrate this in a simple example below, and
will discuss this further in later work.

To see the IBP in action,
 let us apply it to the vacuum polarization
in \vacpoBK:
$$ \eqalign { \Pi =
 \Gamma_2(k_1,k_2) =
{(g\mep)^2\over \fpe} \linv \Tr(T^a T^b) &
 \Big[  \e_1\cdot \e_2\  k_1\cdot k_2
 -     \e_1\cdot  k_2  \e_2\cdot  k_1   \Big]\cr
\loopTiNte{1} & \int_0^{1}  du\   e^{k_1\cdot k_2 G_B(1-u)}
  [\Gdt_B(1-u)]^2       \ .
\cr}
\eqn\vacpoIBP$$
This expression has the remarkable property of being explicitly
transverse.  In usual techniques this property is not
visible until the full set of integrations is complete.
(This is the full result; since the integrand contains
two powers of $\Gbd{12}$, there is no pinch contribution.
Of course this will always be true for a two-point function.)
In fact, \vacpoIBP\ represents precisely
the $(A^\mu)^2$ piece of $F^\mn F_\mn$, which appears as the only
infinite term in the unrenormalized effective action.  In light
of the previous paragraph, it will not surprise the reader
that other infinities, namely the
one-pinch piece of the $(A_\m)^3$ term and the
two-pinch piece of the $(A_\m)^4$ term of the effective action,
reproduce explicitly the remaining
pieces of $F^\mn F_\mn$.  Additionally, since one may perform at
most $N/2$ pinches, there are no infinities beyond $N=4$ in the
effective action.
Thus, even though the complicated process of pinching replaces
the many diagrams of Feynman rules, the IBP and the \BK-type
pinch rules
allow for a clearer
separation of the different types of contributions to the effective
action.  This may prove useful in the analysis of
the divergence structure of more complex theories.

Another interesting feature of this reorganization
is illustrated through the  IBP of \VPFBK:
$$ \eqalign {
 \Pi = -2{(g\mep)^2\over \fpe}& \linv \Tr(T^a T^b)
 \Big[ \ee{1}{2}\kk{1}{2} - \ek{1}{2}\ek{2}{1}  \Big] \cr
\loopTiNte{1}   \int_0^{1}&  du\   e^{k_1\cdot k_2 G_B(1-u)}
  \Big([\Gdt_B(1-u)]^2 - [G_F(1-u)]^2\Big) \   .
\cr}
\eqn\VPFIBP$$
As pointed out by \BnK\refmark{\BKrules,\GSW}, the IBP
allows use of worldline supersymmetry in a
clever way.  Were the system truly worldline supersymmetric, the
effective action would vanish.  Supersymmetry
would require that
 both $x^\m$ and $\psi^\m$ satisfy periodic boundary
conditions, so that $\Gbd{ij}$ and $\Gf{ij}$ would be equal.
It follows that every supersymmetric
amplitude expressed as a function of only $\Gdt_B$ and $G_F$
would vanish
under the formal replacement $\Gbd{ij}\rarr\Gf{ij}$.
However, in
\EAwGBFs\ the only dependence on boundary conditions is
hidden in the Green functions themselves; the functional
dependence on the Green functions is the same in all cases.
As a result,
even when $x^\m$ and $\psi^\m$ have different boundary conditions
the replacement $\Gbd{ij}\rarr \Gf{ij}$ everywhere
in the improved kinematic factor  (and use of momentum
conservation) leads to a complete cancellation.
In particular, the result of \VPFIBP\ has this property.
This trick can be used as a check on the algebra of the IBP.

To find the vacuum polarization for a vector boson loop, follow
the rules in section 4. Specifically, take eq.~\VPFIBP,
replace $(\Gf{21})^2=-\Gf{21}\Gf{12}$ by $+8$,
multiply the term with
$(\Gbd{21})^2$ by 2, and multiply the entire
expression by $-\fourth$:
$$ \eqalign {
 \Pi = {(g\mep)^2\over \fpe}& \linv \Tr(T^a T^b)
 \Big[ \ee{1}{2}\kk{1}{2} - \ek{1}{2}\ek{2}{1}  \Big] \cr
 \loopTiNte{1}   &\int_0^{1}  da\   e^{Tk_1\cdot k_2 (a-a^2)}
  \Big((1-2a)^2 - 4\Big) \cr
 = {(g\mep)^2\over \fpe}& N_c \delta^{ab}
 \Big[ \ee{1}{2}\kk{1}{2} - \ek{1}{2}\ek{2}{1}  \Big] \cr
 \bigg[\loopTiNte{1}   &\int_0^{1}  da\
  \Big((1-2a)^2 - 4\Big) \ + {\rm finite.}\bigg]
\cr}
\eqn\VPVIBP$$
The reader may easily
check that the same result is obtained by integrating
\VPVBK\ by parts, and that the divergent
term yields the usual
$11/3$ associated with the Yang-Mills beta function.

\chapter {Colorful Comments}

It is often desirable to
use a formulation, referred to as ``color-ordered'',
in which only group matrices in the fundamental
representation appear;
the usefulness of this approach
for scattering amplitudes
is detailed in the literature.\refmark{\treelevl,\MPX,\BKcolor}
In particular,
the utility of computing color-ordered tree-level partial amplitudes
using color-ordered Feynman
diagrams was emphasized by
Mangano, Parke and Xu\refmark\MPX.  A study of color-ordering
in loop graphs was performed by \BnK\refmark\BKcolor,
using the techniques of open-string theory, in which
these color traces, known as  Chan-Paton factors\refmark\ChanPat,
appear automatically.

To arrive at such a formulation
in the language of this paper,
 one should write the effective
action as a product of parallel or antiparallel Wilson loops.
Since in $U(N_c)$ the $U(1)$ photon decouples from the $SU(N_c)$ gluons,
one-loop amplitudes for $SU(N_c)$ can be calculated using
$U(N_c)$\refmark\treelevl; working with
the full unitary group allows the use of a number of useful
tricks.\refmark{\treelevl, \BKcolor}
If the particle in the loop
lies in the adjoint representation of
$U(N_c)$, one may consider it as a sort of ``bound state'' of a
fundamental $N_c$ and an antifundamental $\bar {N_c}$
representation; some of the external
vector bosons couple to the $N_c$ while others
couple, independently, to the $\bar {N_c}$.  For a scalar particle,
the effective action is
$$ \eqalign {\Gamma[A]  = \tauint\ \cnst           &\int \Dx\
 \exp\Big[-\taupint (\otuE\xdt^2)\Big]\ \cr
\linv\Tr_{N_c}\ \exp\Big[\taupint& (igA\cdot\xdt)\Big]\
\linv\Tr_{N_c}\
\bigg\{\exp\Big[\taupint (igA\cdot\xdt)\Big]\bigg\}^{\dag}\ ,
\cr }  \eqn\twotrace $$
where the gauge field is a matrix in the fundamental representation.
The first trace is path-ordered, while the second is anti-path-ordered.
In such an expression it becomes immediately obvious that one
expects contributions with one or two group traces at the one
loop-level, as is well-known to
those familiar with the double line formalism
of 't~Hooft\refmark\thooft\ or with open string
theory.\refmark\BKcolor\
Rewriting \EAwGBFs\ in this form changes only
the trace structure: letting $X^a$($T^a$) be the group matrices
in the adjoint (fundamental) representation, we replace
$$ \Tr(X^{a_N}\cdots X^{a_1}) \rarr \sum_{m=1}^{N}
   (-1)^m\Tr(T^{b_{N-m}}\cdots T^{b_1})
\Tr(T^{c_1}\cdots T^{c_m})  \eqn\newtrace $$
where $t_{b_{i+1}}>t_{b_i}$ and $t_{c_{j+1}}>t_{c_j}$.
Thus we divide the gluons into two sets, writing down a path-ordered
trace for one and an anti-path-ordered trace for the other, and
sum over all sets and all orderings.
If $m=0$ or $N$ the trace of the unit matrix yields a factor
of $N_c$. Notice that for $N=2$ the
traces with $m=0$ and $m=2$ are equal, while the case
$N=4$, $m=2$ appears twice in this sum since it is invariant
under proper-time-reversal; this accounts for the factors of
two which appear for these traces in the \BK\ rules.\refmark\BKrules\

Each color trace in \newtrace\ is internally path-ordered,
but operators in {\it different}
traces may be integrated past each other without altering the
color structure.  As a result, surface terms from the IBP
and the operator $\O_{i,j}$ (eq.~\AAterm)
only appear for gluons lying adjacent to each
other in the same color trace; we must therefore only pinch
gluons in the {\it same} trace.
Again this is in agreement with the \BK\ rules.\refmark\BKrules
(For vector particles, the rules for $G_F$ chains
are unaffected by changes in the organization of
color; for a chain to contribute it must still
be path-ordered as in \ordereda\ or \orderedb.)

It may have occurred to the reader educated in string theory
that although I
treated color using a Wilson-loop formalism related to the open
string, I might have introduced color via the use of internal
currents as in the closed string.
This has been discussed in the literature.\refmark\Cascolor\
Such a treatment can easily be implemented,
and rules can be derived using an approach
very similar to that of \BnK\refmark\BKrules;
however this is somewhat more complicated than the technique
used in this paper.

\chapter{Some Extensions}

There are a number of additional theories that are simple to
construct.  For example, to study massive scalars or
spinors in a background gauge field,  add a mass term to
the particle Lagrangian, as in eq.~\freepart:
$$ L \rarr L - \half\EB m^2   \eqn\addmass$$
where $\EB$ is the einbein, and I work in Minkowski spacetime.
{}From the point of view of one-dimensional
general relativity, this is just a cosmological constant.
In the gauge $\EB=2$,
 the scalar effective action becomes
$$\eqalign {\Gamma[A]
 =& - \tauint\ \cnst \int \Dx\
 \exp\Big[-\taupint
(\fourth\xdt^2\  - m^2 - i gA\cdot\xdt)\Big]\ \cr
 =& - \tauint\ \cnst\ e^{+m^2T} \int \Dx\
 \exp\Big[-\taupint
(\fourth\xdt^2\  - i  gA\cdot\xdt)\Big]\ . \cr }
\eqn\EAwmass $$
Thus the effect is merely to add a factor of $e^{+m^2T}$
to the integrand of the integral over T.   Exactly the same factor
occurs for massive spinors.  In Euclidean spacetime the
factor is $e^{-m^2T}$, which illustrates the decoupling of
particles as $m\rarr\infty$.

Another straightforward modification
is the inclusion of background scalars.
Consider the theory
$$ \Lgr = \half(\del\phi)^2 - V(\phi)  \eqn\scalLgr  $$
The one-loop particle Lagrangian of a scalar particle in a background
scalar field can be found
by letting $\phi = \Phi + \d\phi$, where $\d\phi$ is a quantum
fluctuation around the classical field $\Phi$, and keeping only
the terms quadratic in $\d\phi$.
$$L = \otuE\xdt^2 - \half\EB V''(\Phi) . \eqn\sslgr $$
A prime denotes a derivative with respect to $\Phi$.  Notice
that mass terms for the scalar arise correctly from this formula.

Spinors interact with this field in a slightly more
complex way; the Yukawa interaction $h\Phi\bar{\Psi}\Psi$
is easily incorporated in analogy to eq.~\effactFG:
$$ \eqalign{\Gamma[A]
& = \log\ \Big[\det\ \big( i\Dslash-h\Phi\big)\Big]  \cr
& = \half \log\
\Big[\det\ \big( i\Dslash-h\Phi\big) \big(-i\Dslash-h\Phi\big)\Big]  \cr
& = \half \log\ \Big[\det\ \big(\Dslash^2{\bf 1}\
- ih\Dslash\Phi + h^2\Phi^2
\big)\Big]\
\cr } \eqn\EAFwscal $$
The associated spinor particle has Lagrangian
$$L= \otuE\xdt^2 + \half\psi\dot\psi - h^2\Phi^2 +
ih\psi^\m D_\m\Phi\ . \eqn\fslgr $$
Notice that the one-scalar vertex operator for $\Phi=e^{ik\cdot x}$
is $\V_\Phi=-ih (ik\cdot\psi)e^{ik\cdot x}$, as in string theory.
If
we let the scalar field have a vacuum expectation value $v$, and let
$\Phi'=\Phi-v$, then \fslgr\ becomes
$$L= \otuE\xdt^2 + \half\psi\dot\psi -  (hv)^2 - 2h^2v\Phi'
-h^2\Phi'^2 +
ih\psi^\m\del_\m\Phi'\ . \eqn\fslgrvev $$
Of course the particle picks up a mass $m_\Psi=hv$, and the scalar
vertex operator becomes
$\V_\Phi=-ih (ik\cdot\psi-2im_\Psi)e^{ik\cdot x}$.

More interesting is the interaction of a vector boson
with a scalar.  At this point we should remember that
a single background scalar can change the particle in the loop from
a vector into a scalar!  We must therefore build
a theory which consistently
describes a particle that can be either
scalar or vector.  Again string theory is a guide; simply
use dimensional reduction.
Extend the vector theory of section 4
to a fifth dimension (add fields $x^4,\psi_\pm^4$)
but insist that the fifth component of all momenta of all particles
or fields must vanish.
  Since the momentum of the particle must lie
in the usual spacetime, a polarization vector pointing solely in
the $x^4$ direction will always satisfy the physical condition
 $\e\cdot k=0 $; thus the particle's new physical mode
is a Lorentz scalar, while its others are unchanged.  In short, we
have a theory of gauge bosons and a Higgs boson in the adjoint
representation.

The reduction of \backyang\ from five to four dimensions, with
$\Phi\equiv A_4$ and $\phi\equiv Q_4$, is
$$ \eqalign {
 S = \int d^4x\ \{Q^{a\m} [(D^2 + g^2\Phi\Phi)^{ab}g_{\m\n} - g
&(F^c_{\r\s}J^{\r\s})_{\m\n}f^{cab}] Q^{b\n} \cr
+\ gQ^{a\m} (D_\m\Phi)^c\phi^b f^{abc}
 -\ &g\phi^a (D_\m\Phi)^cQ^{b\m} f^{abc} \cr
-\  \phi^a [(D^2 + g^2\Phi\Phi)^{ab}\phi^b
 +\ & \bar{\omega}^a (D^2 + g^2\Phi\Phi)^{ab} \omega^b \cr
+ {\rm order}(Q^3,Q^4,DQ\phi^2,\bar{\omega}Q\omega,etc.)&\} \ .\cr }
  \eqn\dimredYM $$
This formula stems from the gauge $D^\m Q_\m+ig[\Phi,\phi]=0$,
called background 't~Hooft-Feynman gauge.
Notice that this gauge contains a new, gauge dependent $\Phi^2\phi^2$
interaction, different from the  $\VEV{\Phi^2}\phi^2$ interaction
present in usual 't~Hooft-Feynman gauge\refmark\tHgauge, in which
$\del^\m Q_\m + ig[\VEV{\Phi},\phi] = 0$.
It is clear from \dimredYM\ that if $\Phi$ acquires a
vacuum expectation value
the gluons, ghosts, and Goldstone bosons
associated with spontaneously broken generators have
the same mass matrix:
$$ (M^2)^{ab} =
g^2\VEV{\Phi\Phi}^{ab} = g^2f^{ace}f^{bde}\VEV{\Phi^c\Phi^d}   \ .
\eqn\massmtrx$$
It is straightforward to add in the symmetry breaking potential
for the Higgs boson, and to extend this approach to Higgs bosons in
other representations.

The particle Hamiltonian for this theory is
$$  H = (p_\m-gA_\m)^2 - (p_4-g\Phi)^2
-  ig\psi^\mu F_{\mu\nu}\psi^\nu\
+ 2ig\psi^\mu D_\m\Phi\psi^4 \ ;
    \eqn\ssvctham $$
when $p_4$ is set to zero, the resulting Lagrangian is
$$  \eqalign{ L = \otuE \xdt^2\ & +
\psp\cdot\dot\psm - \psp^4\dot\psm^4
 -g^2\Phi^2 -igA^\m \xdt_\m \cr
&+  ig\psi^\mu F_{\mu\nu}\psi^\nu\
+ 2ig\psi^\mu D_\m\Phi\psi^4 \ .
 }   \eqn\ssvctlgr $$
The last term is the one that turns a scalar in the loop into
a  vector, and {\it vice versa}.
When $\VEV{\Phi}$ is non-zero the mass matrix of \massmtrx\
is clearly generated.  To add in a Higgs potential $V(\Phi)$,
use
$$L \rarr L+V''(\Phi)(\psp^4\psm^4) \ ; \eqn\VSSB $$
the oscillator potential for $\psi^4$ assures that
of the physical states only $\psp^4\ket{0}$, the scalar, will
feel the potential.  This sort of theory can be used --- perhaps
profitably --- for calculations in the standard model; a set
of rules is in preparation.

Particles in background gravity may also be treated in this way.
Consider a
theory of a scalar boson in a background metric $G_\mn$:
$$ L = \otuE  G_\mn\xdt^\m\xdt^\n
- \Eotu m^2 \ .
\eqn\backgrav $$
This is generally covariant with respect to both
worldline and spacetime coordinate redefinitions.
One may extend this theory to particles with spin.
The relevant Lagrangians were again written down by Brink,
Di Vecchia and Howe\refmark\BdVH, and I shall not repeat them
here.
However, quantization of such a Lagrangian is subtle.\refmark\GJ\
The technique for constructing internal gravitons also appears
in reference \BdVH: instead of one complex set of worldline
fermions,  use two.  Define a particle with an N=4 worldline
supersymmetry, described by coordinates
($x^\m,\psi^\m_\pm,\chi^\m_\pm$).
The allowed states can be written down as in \thestates; projections
onto odd $\psi$ and $\chi$ number and onto states which are even
under $\psi\rarr\chi$
leaves a rank-two symmetric tensor as the propagating
modes of the theory.
While not particularly elegant, this example illustrates
that it is
straightforward to  construct a tensor of any arbitrary rank and
symmetry.  It may be hoped that useful rules
can be obtained from this theory as well.

Finally, I should point out that every theory described in this
paper is part of the mode expansion of a string in a background
string field.\refmark\backstring\
The possible connection of this construction
to the \BK\ rules was noted by Bern and
Dunbar.\refmark\BDmap\

\chapter{Conclusion}

In this paper, I have shown that it is possible to construct one-loop
effective actions perturbatively without the use of Feynman diagrams,
and with a method that has certain
conceptual and practical advantages over
the standard technique.  By viewing a one-loop computation as a
system of a particle (or superparticle) in a background field, one
can construct formulas and rules
valid to all orders in the background field
which closely match the string-derived
rules of \BnK\ for gauge theory.  It is now evident that one reason
for the simplicity of the \BK\ rules compared to Feynman diagrams
is that string theory is a
first-quantized system; the ease of
one-dimensional as opposed to four-dimensional calculations
is clearly demonstrated both in this paper and in the work of \BnK.
The formalism developed in this paper represents a
technical and conceptual shift
away from the standard techniques of path integral perturbative
field theory and back to basic quantum mechanics
and the background field method.

\medskip
{\titlestyle{\twelvepoint\bf{Appendix: Conventions}}}

In this paper I have used conventions which are appropriate for
particles and Wilson loops and which generate expressions
that are simple to compare with those of
Feynman diagrams.  Unfortunately
they are not the most convenient from all points of view, and indeed
\BnK\ have chosen a very different set of conventions.  It is
straightforward to convert from one to the other,
and in this appendix I explain how to do so.

First, let me review my conventions.  I use
$$\eqalign {
g_\mn = {\rm diag}\{+ - - -\}\ \ &; \ \ \
\Tr[T^aT^b] = \half \delta^{ab} \ ; \cr
D_\m = \del_\m - igA_\m \ \ &; \
\ \ gF_\mn = i[D_\m,D_\n] \cr } \eqn\conventA $$
and for Grassmann integrations
$$ \int d\ta d\tab\  \tab\ta = 1 \ . \eqn\conventB $$

To convert my expressions to those of \BnK:\refmark\BKPascos\
\point
 Reverse the order of the color trace.
\point
 Write the Grassmann integral of \EAwGBFs\ as $\int d\tab_i d\ta_i$
(but keep eq.~\conventB.
\point
 Replace $\Gdt_B$ with $-\Gdt_B$.
\point
 Divide all $\Gdt_B$ and $G_F$ functions by 2.
\point
 Multiply all group matrices by $\sqrt{2}$.
\point
 Account for these factors of two by multiplying the entire
amplitude by $2^{N/2}$.

As a result,
\point
 The improved kinematic factor vanishes under $\Gdt_B\rarr -G_F$.
\point
 Pinches at a vertex with gluons $j$ and $i$, $j$ the most clockwise,
 result in the replacement
 $k_i\cdot k_j \Gbd{ji}(\Gf{ji}) \rarr +(-)\half$.

\medskip

{\titlestyle{\twelvepoint\bf{Acknowledgements}}}

I have benefited enormously from discussions with a number
of physicists; their ideas and insights appear throughout
this paper.  Z.~Bern and D.~A.~Kosower answered
many questions and
helped me to understand the relation of their rules
to usual field theory concepts.
I especially thank Z.~Bern for explaining to me the role
of Schwinger proper time and for discussions
on gauges, tree diagrams, and the integration-by-parts procedure.
I thank L.~J.~Dixon for explaining many aspects
of string theory, and particularly
for important discussions about the mode expansion of the string.
R.~Kallosh clarified certain issues concerning
 supersymmetry and the background
field method, and also pointed
me toward the work of Brink, Di Vecchia and Howe.
 D.~C.~Lewellen advised looking at
first-quantized field theory and suggested several useful papers.
In addition to helping me with the many subtleties of string theory,
M.~E.~Peskin repeatedly pointed out the value of Wilson loops
in gauge theory, and made useful
observations regarding the integration-by-parts
procedure and manifest gauge invariance.
I also had useful discussions with
S.~Ben-Menachem, A.~W.~Peet,
Y.~Shadmi, L.~Susskind, L.~Thorlacius and B.~J.~Warr.

\refout
\end